\documentclass[twocolumn,pre,twoside,superscriptaddress,floatfix]{revtex4-1}

\usepackage{adjustbox}
\usepackage{amsmath}
\usepackage{amssymb}
\usepackage{graphicx}
\usepackage{dcolumn}
\usepackage{bm}
\usepackage{color}
\usepackage[T1]{fontenc}

\def\epsilon{\varepsilon}

\def\theta{\vartheta}
\def\rho{\varrho}
\def\Int#1#2#3{\int_{#1}\!\mathrm{d}^{#2}{#3}\;}

\def\vec#1{\pmb{#1}}

\definecolor{gruen}{rgb}{0.0,0.7,0.2}
\definecolor{Magenta}{RGB}{186,085,211}
\definecolor{orange}{RGB}{238,154,073}
 \def\modifiedRed#1{\textcolor{black}{#1}}
 \def\modifiedBlue#1{\textcolor{black}{#1}}
 \def\modifiedGreen#1{\textcolor{black}{#1}}
 \def\modifiedMagenta#1{\textcolor{black}{#1}}
 \def\MODIFIED#1{\textcolor{black}{#1}}
 \def\MODIFIEDtwo#1{\textcolor{black}{#1}}

\graphicspath{{Figs/}}

\begin{document}


\title{Interface structures in ionic liquid crystals}

\author{Hendrik Bartsch}
\email{hbartsch@is.mpg.de}
\affiliation
{
   Max-Planck-Institut f\"ur Intelligente Systeme,\\ 
   Heisenbergstr.\ 3,
   70569 Stuttgart,
   Germany
}
\affiliation
{
   Institut f\"ur Theoretische Physik IV,
   Universit\"at Stuttgart,
   Pfaffenwaldring 57,
   70569 Stuttgart,
   Germany
}
\author{Markus Bier}
\email{bier@is.mpg.de}
\affiliation
{
   Max-Planck-Institut f\"ur Intelligente Systeme,\\ 
   Heisenbergstr.\ 3,
   70569 Stuttgart,
   Germany
}
\affiliation   
{
   Institut f\"ur Theoretische Physik IV,
   Universit\"at Stuttgart,
   Pfaffenwaldring 57,
   70569 Stuttgart,
   Germany
}
\affiliation
{
   Fakult\"{a}t Angewandte Natur- und Geisteswissenschaften,
   Hochschule f\"{u}r angewandte Wissenschaften W\"{u}rzburg-Schweinfurt,
   Ignaz-Sch\"{o}n-Str.\ 11, 97421 Schweinfurt, Germany
}
\author{S.\ Dietrich}
\affiliation
{
   Max-Planck-Institut f\"ur Intelligente Systeme,\\ 
   Heisenbergstr.\ 3,
   70569 Stuttgart,
   Germany
}
\affiliation
{
   Institut f\"ur Theoretische Physik IV,
   Universit\"at Stuttgart,
   Pfaffenwaldring 57,
   70569 Stuttgart,
   Germany
}

\date{ \today }

\begin{abstract}
Ionic liquid crystals (ILCs) are anisotropic mesogenic molecules which additionally carry charges.
This combination gives rise to a complex interplay of the underlying (anisotropic)
contributions to the pair interactions\modifiedRed{. It} promises interesting and distinctive structural
and orientational properties to arise in systems of ILCs, combining properties of liquid crystals
and ionic liquids.
While previous theoretical studies have focused on the phase behavior of ILCs and the structure of the
respective bulk phases, in \modifiedRed{the present study we provide} new results, obtained within density
functional theory, concerning (planar) free interfaces between an isotropic liquid $L$ and two types of
smectic-A phases ($S_A$ or $S_{AW}$). We discuss the structural and orientational properties of these
interfaces in terms of the packing fraction profile $\eta(\vec{r})$ and the orientational order parameter
profile $S_2(\vec{r})$ \modifiedRed{concerning the tilt angle} $\alpha$ between the (bulk) smectic layer
normal and the interface normal. The asymptotic decay of $\eta(\vec{r})$ and \modifiedRed{of} $S_2(\vec{r})$
towards their values in the isotropic bulk \modifiedRed{is discussed, too.}
\end{abstract}

\maketitle

\section{\label{sec:intro}Introduction}

Ionic liquid crystals (ILCs) are pure ionic systems, solely composed of cations \modifiedRed{($+$)} and
anions \modifiedRed{($-$)}.
\MODIFIED{Moreover, at least one of the ion species is characterized by a highly anisotropic 
molecular shape~\cite{Binnemans2005}. This anisotropic shape is typically due to long alkyl-chains
\MODIFIEDtwo{which} are attached to charged moieties.
Although the alkyl-chains exhibit a rather strong flexibility, due to 
microphase segregation of the charged parts and of the alkyl-chains, liquid-crystalline phases
are indeed observable \MODIFIEDtwo{among} ILCs~\cite{Bowlas1996,Binnemans2005,Goossens2016}.}
In the past decades various types of ILCs have been synthesized~\cite{Binnemans2005,Goossens2016}.
Different combinations of, e.g., (charged) imidazolium rings and alkyl-chains allow one to
tune not only the length of the ionic mesogenes but also the location of their charges, i.e.,
the intra-molecular charge distribution. Thereby one is able to promote distinctive properties of ILCs,
for instance, a high thermal and high electrochemical stability, which might be beneficial for technological
applications~\cite{Gordon_et_al1998,Lee_et_al2003,Binnemans2005,Ster_et_al2007,Goossens2016}.
\MODIFIED{(We note, that here the term ``mesogene'' refers to any kind of molecule \MODIFIEDtwo{which}
gives rise to the formation of mesophases, irrespective of the underlying microscopic mechanism.
\MODIFIEDtwo{Accordingly}, the aforementioned anisotropic molecules, which form mesophases via
microphase segregation, are considered to be mesogenes.)}

\MODIFIED{A specific example of an ILC system, which has been studied, e.g., in
Refs.~\cite{Yamanaka_et_al2005,Wang_et_al2012}, is composed of cations with long alkyl-chains
attached (1-dodecyl-3-methylimidazolium) and significantly smaller anions (iodide).
For such an ILC system, one observes a liquid crystalline structure, in particular the
smectic-A phase $S_A$. (The $S_A$ phase is characterized by layers of particles which are well
aligned with the layer normal and the layer spacing is of the size of the particle length.)
The layer structure of the large cations leads to a locally increased concentration of anions
in between the layers of cations~\cite{Yamanaka_et_al2005}. Thereby, the nanostructure of the
cations gives rise to ``pathways'' for the anions, which increase the conductivity measurable
in the direction parallel to the layers.}
Therefore this particular type of an ILC system is a promising candidate for technological
applications, e.g., as electrolyte in dye-sensitized solar cells
(DSSCs)~\cite{Yamanaka_et_al2005,Yamanaka_et_al2007}.

While the complexity of the underlying interactions gives rise to these interesting properties of ILCs,
it is at the same time very challenging to study these systems within theory or simulations.
\MODIFIED{Previous theoretical studies~\cite{Kondrat_et_al2010,Bartsch2017} \MODIFIEDtwo{of} ILC systems
\MODIFIEDtwo{have been} able to reduce this complexity by considering a simplified description of
ILC systems, which incorporates, \MODIFIEDtwo{however}, the generic properties of ILCs.
They rely on an effective one-species description in which one of the
ion species (referred to as counterions) is not accounted for explicitly, but is incorporated as a continuous
background, giving rise to \MODIFIEDtwo{the} screening of the coions.
\MODIFIEDtwo{On the contrary, the coions} are modeled as ellipsoidal 
particles. Thus, the anisotropic molecular shape, \MODIFIEDtwo{which} gives rise to the formation of
mesophases, and the (screened) electrostatic interaction are both incorporated by this approach.
Of course, this is a simplified representation of any realistic ionic liquid crystalline system.
\MODIFIEDtwo{However, it allows one} to study the interplay of the two key features, i.e.,
an anisotropic molecular shape and the presence of charges, which are \MODIFIEDtwo{omnipresent} in ILC systems.
Yet it should be noted, that ILC systems \MODIFIEDtwo{exhibiting} a significant difference in size
of the cations and \MODIFIEDtwo{of the} anions (e.g., the aforementioned example of
1-dodecyl-3-methylimidazolium) might be candidates \MODIFIEDtwo{which} come closest to the present
theoretical representation of ILCs, as the size difference rationalizes in parts the idea of
structureless point-like counterions.}

As a first step, \MODIFIED{such a model} allows one to study the phase behavior of ILC systems and thereby
\modifiedRed{to gain insight about} how molecular properties, e.g., the aspect-ratio or the charge
distribution of the molecules, affect the phase behavior of such types of ILCs.
A comprehensive understanding of the relation between the underlying molecular properties and the resulting
phase behavior is \modifiedRed{inter alia,} necessary for a systematic synthesis of ILCs,
which should meet specific material properties. Furthermore, theoretical \modifiedRed{guidance}
is beneficial for finding and exploring novel \modifiedRed{materials properties which might occur}
in ILC systems. For instance, in Ref.~\cite{Bartsch2017} a new smectic-A structure ($S_{AW}$) has been
observed, which \modifiedRed{exhibits} an alternating layer structure. In between layers of
\modifiedGreen{elongated} particles, which prefer to be \modifiedGreen{oriented} parallel to the
layer normal, like in the ordinary $S_A$ phase, one observes secondary
layers in which the particles prefer to be \modifiedRed{oriented} perpendicular to it.
Due to this alternating structure the layer spacing of this new $S_{AW}$ phase is significantly
\textit{w}ider compared to the ordinary $S_A$ phase.
The $S_{AW}$ structure is stabilized by charges \modifiedRed{which} are located at the tips of
the molecules. This shows in an exemplary way  how the combination of liquid-crystalline behavior
and electrostatics can lead to \modifiedRed{an} interesting and novel phenomenology.

The aim of the present investigation is to extend the analysis by studying spatially
inhomogeneous systems of ILCs. This is done by investigating how the structural and
orientational properties of ILC systems are affected by the presence of a free interface between 
coexisting bulk states. Both smectic-A phases, $S_A$ and $S_{AW}$, observed in Ref.~\cite{Bartsch2017}
can be in coexistence with the isotropic liquid phase $L$. This is of intrinsic interest, because it
allows one to investigate interfaces which interpolate between a structured and orientationally ordered
(i.e., smectic) phase and an isotropic, homogeneous, and thus structure-less, fluid phase.
\MODIFIED{In particular, the transition in the structural and \MODIFIEDtwo{in the} orientational order
allows one to study the interplay of both properties while they build up at the interface.
Although there are theoretical analyses~\cite{Mederos1992,Somoza1995,Martinez-Raton1998,DeLasHeras2005,Wolfsheimer2006,Reich2007,Praetorius_et_al2013}
\MODIFIEDtwo{concerning} related types of free interfaces, in these studies the \MODIFIEDtwo{constituent}
particles are \MODIFIEDtwo{plain} liquid crystals without any charges.
\MODIFIEDtwo{On the other hand, there is a} vast number of theoretical studies on ionic fluids.
The thermodynamic behavior~\cite{Fisher1993,Fisher1994,Luijten2002,Kobelev2002} as well as
the structure~\cite{Stillinger1_1968,Stillinger2_1968,Lovett1968,Mitchell1977,Harnau2000} of these
types of fluids, in which long-ranged Coulomb interactions are present, have been intensively studied.
However, ionic systems are often analyzed assuming a simple geometry of the \MODIFIEDtwo{particles},
\MODIFIEDtwo{such as} a spherical shape of the particles like in the restricted primitive
model~\cite{Stell1976,Gillan1983,Dickman1999}.
In \MODIFIEDtwo{this} regard, the present \MODIFIEDtwo{study} attempts to \MODIFIEDtwo{analyze} the 
\MODIFIEDtwo{aforementioned} type of interface between an isotropic and a smectic phase by
accounting for an anisotropic particle shape combined with the presence of charges.}

Moreover, different orientations between the interface normal and the smectic layer normal are possible.
In this context, an interesting question addresses the equilibrium tilt angle between the interface and
the smectic layer normal. This angle may provide insight into nucleation and growth phenomena which are
affected by the dependence of the interfacial tension on the orientation of the considered
structure~\cite{Wulff1901,Blanc2001}.

The present study is structured as follows:
In Sec.~\ref{sec:theory} the model and the employed density functional theory approach
are presented. Our results for the interfaces between the isotropic liquid $L$ and the considered
smectic-A phases $S_A$ or $S_{AW}$ are discussed in Sec.~\ref{sec:results}.
Finally, in Sec.~\ref{sec:summary} we summarize the results and draw our conclusions.

\section{\label{sec:theory}Model and methods}

This section presents in detail the molecular model of ILCs as employed here.
In particular, we discuss the intermolecular pair potential, which can be applied to a wide range of
ionic and liquid crystalline materials due to its flexibility provided by a large set of parameters.

This model is studied by (classical) density functional theory (DFT), which will be applied to
\modifiedRed{spatially} inhomogeneous systems, in particular free interfaces formed between coexisting
bulk phases. The methodological and technical details of the present DFT approach are described in
Sec.~\ref{sec:theory:DFT}.
\begin{figure}[!t]
 \includegraphics[width=0.45\textwidth]{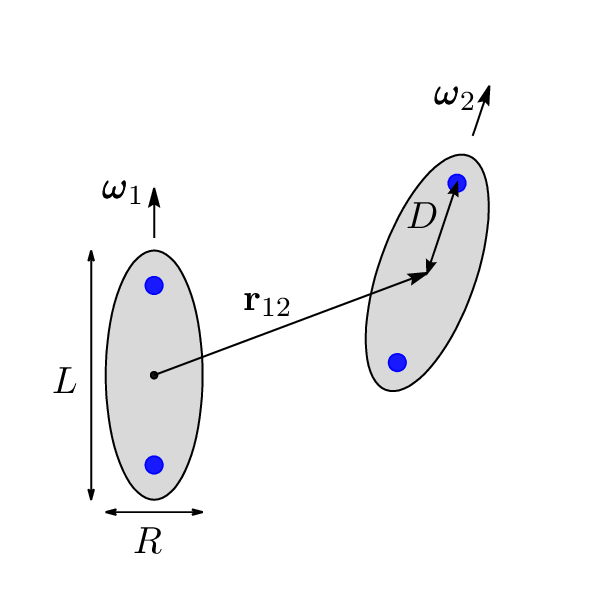}
  \caption{
  Cross-sectional view of two ILC molecules
  in the plane spanned by the orientations $\vec\omega_i,~i=1,2$,
  of their long axis.
  The particles are treated as rigid prolate ellipsoids,
  characterized by their length-to-breadth ratio \modifiedRed{$L/R\geq1$}.
  Their orientations are fully described by the direction of
  their long axis $\vec\omega_i, \modifiedRed{i=1,2}$; 
  $\vec{r}_{12}$ is the center-to-center distance vector.
  The charges of the ILC molecules (blue dots) are
  located on the long axis
  at a distance $D$ from their geometrical center.
  The counterions are not modeled explicitly,
  but they are implicitly accounted for in terms of a background,
  giving rise to the screening of the charges of the ILC molecules. 
  }
 \label{fig:ellipsoids}
\end{figure}
%

\subsection{\label{sec:theory:model}Molecular model and pair potential}
We consider a coarse-grained description of the ILC molecules as rigid prolate ellipsoids of
length-to-breadth ratio \modifiedRed{$L/R\geq1$} (see Fig.~\ref{fig:ellipsoids}).
Thus, the orientation of a molecule is fully described by the direction $\vec\omega(\phi,\theta)$
of its long axis, where $\phi$ and $\theta$ denote the azimuthal and polar angle, respectively.

The two-body interaction potential consists of a hard core repulsive and an additional contribution
$U_\text{GB}+U_\text{es}$ beyond the contact distance $R\sigma$, the sum of which can
be attractive or repulsive:
\begin{equation}
   U= 
   \begin{cases}
      \infty
      &, 
      |\vec{r}_{12}|    < R\sigma(\vec{\hat r}_{12},\vec\omega_1,\vec\omega_2) \\
      \begin{split}
       U_\text{GB}(\vec{r}_{12},\vec\omega_1,\vec\omega_2)+\\
       U_\text{es}(\vec{r}_{12},\vec\omega_1,\vec\omega_2)
      \end{split}
      &,
      |\vec{r}_{12}| \geq R\sigma(\vec{\hat r}_{12},\vec\omega_1,\vec\omega_2),
   \end{cases}
\label{eq:Pairpot}
\end{equation}
where $\vec{r}_{12}:=\vec{r}_2-\vec{r}_1$ denotes the center-to-center distance vector
between the two particles labeled as 1 and 2, and $\vec\omega_i$, $i=1,2$, are their orientations
\modifiedRed{with $|\vec\omega_i|=1$}.
The contact distance $R\sigma(\vec{\hat r}_{12},\vec\omega_1,\vec\omega_2)$ depends on the orientations
of both particles and \modifiedRed{on} the direction of the center-to-center distance vector,
which is expressed by the unit vector $\vec{\hat r}_{12}:=\vec{r}_{12}/|\vec{r}_{12}|$.
In Eq.~(\ref{eq:Pairpot}), we \modifiedRed{have} subdivided the contributions beyond the contact distance
$|\vec{r}_{12}|\geq R\sigma$ into two parts:
$U_\text{GB}(\vec{r}_{12},\vec\omega_1,\vec\omega_2)$ is the well-known Gay-Berne
potential~\cite{Berne1972,Gay_Berne1981}, which incorporates an attractive van der Waals-like interaction
between molecules and which can be understood as a generalization of the Lennard-Jones pair potential
\modifiedRed{between spherical particles} to ellipsoidal particles:
\begin{equation}
 \begin{split}
  &    U_\text{GB}(\vec     {r}_{12},\vec\omega_1,\vec\omega_2)
  =      4\epsilon(\vec{\hat r}_{12},\vec\omega_1,\vec\omega_2)\\
  & \times \left[  \left(1+\frac{|\vec{r}_{12}|}{R}-\sigma(\vec{\hat r}_{12},\vec\omega_1,\vec\omega_2) \right)^{-12} \right.\\
  &        \left.-~\left(1+\frac{|\vec{r}_{12}|}{R}-\sigma(\vec{\hat r}_{12},\vec\omega_1,\vec\omega_2) \right)^{-6}  \right]\\
 \end{split}
 \label{eq:Pairpot_GB}
\end{equation}
with
\begin{equation}
 \begin{split}
    \sigma(\vec{\hat r}_{12},\vec\omega_1,\vec\omega_2)
  & =\left[ 1-\frac{\chi}{2}\left(\frac{(\vec{\hat r}_{12}\cdot(\vec\omega_1+\vec\omega_2))^2}{1+\chi\vec\omega_1\cdot\vec\omega_2}\right.\right.\\
  &  \left.  +              \left.\frac{(\vec{\hat r}_{12}\cdot(\vec\omega_1-\vec\omega_2))^2}{1-\chi\vec\omega_1\cdot\vec\omega_2}\right)\right]\\
 \end{split}
\end{equation}
and
\begin{equation}
 \begin{split}
     \epsilon(\vec{\hat r}_{12},\vec\omega_1,\vec\omega_2)
  & =\epsilon_0\left(1-(\chi\vec\omega_1\cdot\vec\omega_2)^2\right)^{-1/2}\\
  & \times\left[ 1-\frac{\chi'}{2}\left(\frac{(\vec{\hat r}_{12}\cdot(\vec\omega_1+\vec\omega_2))^2}{1+\chi'\vec\omega_1\cdot\vec\omega_2}\right.\right.\\
  & \left.  +               \left.\frac{(\vec{\hat r}_{12}\cdot(\vec\omega_1-\vec\omega_2))^2}{1-\chi'\vec\omega_1\cdot\vec\omega_2}\right)\right].\\
 \end{split}
 \label{eq:Pairpot_GB_epsilon}
\end{equation}
The contact distance $R\sigma(\vec{\hat r}_{12},\vec\omega_1,\vec\omega_2)$ and the direction- and
orientation-dependent interaction strength $\epsilon(\vec{\hat r}_{12},\vec\omega_1,\vec\omega_2)$ are both
parametrically dependent on the length-to-breadth ratio $L/R$ via the auxiliary function 
$\chi=((L/R)^2-1)/((L/R)^2+1)$. Additionally, $\epsilon(\vec{\hat r}_{12},\vec\omega_1,\vec\omega_2)$
can be tuned via $\chi'=((\epsilon_R/\epsilon_L)^{1/2}-1)/((\epsilon_R/\epsilon_L)^{1/2}+1)$,
where $\epsilon_R/\epsilon_L$ is called the anisotropy parameter, defined in terms of the ratio of
$\epsilon_R$, which is the depth of the potential minimum for parallel particles positioned side by side
$(\vec{\hat r}_{12}\cdot\vec\omega_1=\vec{\hat r}_{12}\cdot\vec\omega_2=0)$, and $\epsilon_L$, which is
the depth of the potential minimum for parallel particles positioned end to end
$(\vec{\hat r}_{12}\cdot\vec\omega_1=\vec{\hat r}_{12}\cdot\vec\omega_2=1)$.
The energy scale of the Gay-Berne pair interaction is set by $\epsilon_0$.
Thus, the Gay-Berne pair potential has four independent free parameters:
$\epsilon_0, R, L/R$, and $\epsilon_R/\epsilon_L$. Note that in the case of spherical particles, i.e.,
for $L=R$, the Gay-Berne pair potential (Eq.~(\ref{eq:Pairpot_GB})) reduces to the well-known isotropic
Lennard-Jones pair potential iff, additionally, the Gay-Berne anisotropy parameter equals unity, i.e.,
$\epsilon_R/\epsilon_L=1$, because then $\sigma(\vec{\hat r}_{12},\vec\omega_1,\vec\omega_2)=1$ and
$\epsilon(\vec{\hat r}_{12},\vec\omega_1,\vec\omega_2)=\epsilon_0$.

The second contribution $U_\text{es}(\vec{r}_{12},\vec\omega_1,\vec\omega_2)$ in Eq.~(\ref{eq:Pairpot}) is
the \emph{e}lectro\emph{s}tatic repulsion of ILC molecules. Within the scope of the present study,
the counterions are not modeled explicitly. \modifiedRed{They} will be considered to be much smaller in size
than the ILC molecules such that they can be treated as a continuous background.
On the level of linear response, this background gives rise to the screening of the pure Coulomb potential
between two charged sites on a length scale given by the Debye screening length $\lambda_D$ such, that
the effective electrostatic interaction of the ILC molecules is given by
\begin{equation}
 \begin{split}
      U_\text{es}(\vec     {r}_{12},\vec\omega_1,\vec\omega_2)=~
  &   \gamma\left[\frac{\exp\left(-\frac{|\vec{r}_{12}+D(\vec\omega_1+\vec\omega_2)|}{\lambda_D}\right)}{|\vec{r}_{12}+D(\vec\omega_1+\vec\omega_2)|}\right.\\  
  &   +     \left.\frac{\exp\left(-\frac{|\vec{r}_{12}+D(\vec\omega_1-\vec\omega_2)|}{\lambda_D}\right)}{|\vec{r}_{12}+D(\vec\omega_1-\vec\omega_2)|}\right.\\  
  &   +     \left.\frac{\exp\left(-\frac{|\vec{r}_{12}-D(\vec\omega_1+\vec\omega_2)|}{\lambda_D}\right)}{|\vec{r}_{12}-D(\vec\omega_1+\vec\omega_2)|}\right.\\  
  &   +     \left.\frac{\exp\left(-\frac{|\vec{r}_{12}-D(\vec\omega_1-\vec\omega_2)|}{\lambda_D}\right)}{|\vec{r}_{12}-D(\vec\omega_1-\vec\omega_2)|}\right].  
 \end{split}
\label{eq:PairPot_ES}
\end{equation}
The charges $q$ are located symmetrically on \modifiedRed{the long axis of the ILC molecules}
at a distance $D$ from the geometrical center of the particles (compare Fig.~\ref{fig:ellipsoids}).
\MODIFIED{The prefactor $\gamma=q^2/(4\pi\epsilon)$ of dimension $[\text{energy}]\times[\text{length}]$
characterizes the electrostatic energy scale, where $\epsilon$ denotes the permittivity.}
In principle, the Debye screening length 
\begin{equation}
 \MODIFIED{\lambda_D\propto\sqrt{\frac{T}{\rho_\text{c}}}}
\label{eq:Debyelength}
\end{equation}
is a function of temperature $T$ and of the number density $\rho_\text{c}$ of the counter ions. Thus,
it depends on the thermodynamic state of the fluid. However, in the present model $\lambda_D$ is taken
to be a constant parameter.
In order to compare results, obtained within this model, with data from actual physical systems,
one could measure the value of the Debye screening length experimentally and tune the model
parameter $\lambda_D$ accordingly.

\begin{figure}[!t]
 \includegraphics[width=0.45\textwidth]{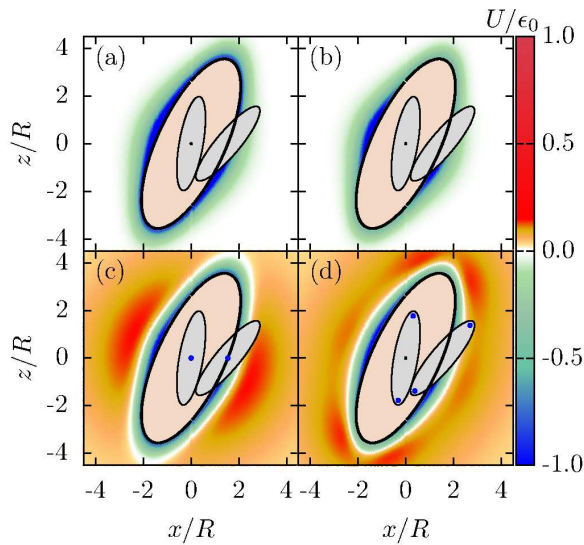}
  \caption{
  Contour-plots of the pair potential $U$ 
  for $|\vec{r}_{12}|\geq R\sigma$ in the $x$-$z$-plane
  for four cases of particles with fixed length-to-breadth ratio $L/R=4$ and fixed orientations.
  In each panel the centers of both particles lie in the plane $y=0$.
  In order to illustrate the orientations of the ellipsoids,
  they have been included in the plots at contact
  with relative direction $\vec{\hat r}_{12}=\vec{\hat x}$.
  The set of points at contact in the $x$-$z$-plane is illustrated by the black curve, and the 
  centers of the particles are shown by small black dots.
  Panel (a): uncharged liquid crystal with $\epsilon_R/\epsilon_L=2$.
  Panel (b): uncharged liquid crystal with $\epsilon_R/\epsilon_L=4$.
  \modifiedRed{With this choice} the anisotropy of the potential is increased slightly.
  Panel (c): ILC with $\epsilon_R/\epsilon_L=2,D/R=0,\lambda_D/R=5,\gamma/(R\epsilon_0)=0.25$.
  Panel (d): ILC with $\epsilon_R/\epsilon_L=2,D/R=1.8,\lambda_D/R=5,\gamma/(R\epsilon_0)=0.25$.
  In (c) and (d) the loci of the charges are indicated as blue dots.
  The salmon-colored area is the excluded volume for given orientations of
  the two particles.
  }
 \label{fig:pairpot}
\end{figure}
In Fig.~\ref{fig:pairpot} we illustrate the full pair potential (Eq.~(\ref{eq:Pairpot}))
beyond the contact distance for certain choices of the parameters.
The two top panels, (a) and (b), show the pure Gay-Berne potential (uncharged liquid crystals), which is
predominantly attractive in the space outside the overlap volume (\modifiedRed{salmon}-colored area).
The shape of this overlap volume changes by varying the particle orientations as well as by changing the
length-to-breadth ratio $L/R$. However, these dependences are not apparent from Fig.~\ref{fig:pairpot},
\modifiedRed{because} $L/R=4$ and the particle orientations $\vec\omega_i$ are kept fixed for all panels.
In panel (b) the anisotropy parameter $\epsilon_R/\epsilon_L=4$ is chosen to be  two times larger than for
panel (a) ($\epsilon_R/\epsilon_L=2$). Thus, the ratio of the well depth at the tails and at the sides is
increased. The two bottom panels, (c) and (d), show the same choices for the Gay-Berne parameters as for
panel (a), but the electrostatic repulsion of the charged groups on the molecules, illustrated by blue dots,
is included ($\gamma/(R\epsilon_0)=0.25$). In panel (c) the loci of the two charges of the particles coincide
at \modifiedRed{their centers} (i.e., $D/R=0$) while in panel (d) they are located near the tips ($D/R=1.8$).
For both cases with charge, the effective interaction range is significantly increased compared with the
uncharged case and is governed by the Debye screening length, chosen as $\lambda_D/R=5$.

\MODIFIED{It is worth mentioning, that the present model cannot be considered \MODIFIEDtwo{as}
a quantitatively valid description of any realistic ionic liquid crystal system.
A screened electrostatic pair interaction of the Yukawa form (Eq.~(\ref{eq:PairPot_ES})) is the
extreme case of the effective pair potential between ions in a (dilute) electrolyte at high
temperatures. Nonetheless, for the purpose of the present theoretical study, which is concerned
with the basic microscopic mechanisms and the generic molecular properties present in ILC systems,
the employed model is appropriate as it incorporates the following key properties of ILCs:
First, a \MODIFIEDtwo{sufficiently} anisotropic shape (prolate) of the particles, i.e.,
they can be considered as (calamitic) mesogenes.
In this context, \MODIFIEDtwo{an assessment} of the bulk phase behavior, depending on the 
length-to-breadth ratio of the particles, is provided in Ref.~\cite{Bartsch2017}. In particular, 
the relevance of a sufficiently anisotropic shape (i.e., $L/R>2$) \MODIFIEDtwo{for observing} genuine
smectic phases is discussed. Second, the ionic properties of ILCs are incorporated such that they
reflect the main feature of ionic fluids, i.e., the \emph{effective} interaction of the ionic compounds
via a \emph{screened} electrostatic pair interaction. 
Although the chosen functional form given by Eq.~(\ref{eq:PairPot_ES}) cannot be considered \MODIFIEDtwo{as}
a quantitatively reliable representation, it still accounts for the fact that the actual ion-ion pair
interaction in an ionic fluid is indeed \MODIFIEDtwo{short-ranged}, rather than \MODIFIEDtwo{long-ranged},
as \MODIFIEDtwo{it is the case} for the bare Coulomb interaction.}

\MODIFIED{In conclusion, Eq.~(\ref{eq:PairPot_ES}) is \MODIFIEDtwo{characterized} by an effective
interaction strength \MODIFIEDtwo{$\gamma/R$} (which will be numerically expressed as the relative
interaction strength $\gamma/(R\epsilon_0)$ compared to the interaction strength $\epsilon_0$ of the
Gay-Berne potential), an effective interaction range $\lambda_D$, and an effective location $D$ of
the charge sites \MODIFIEDtwo{inside} the coions.
In order to represent specific ILC molecules by a particular set of parameters
of the present model, one would tune the independent model parameters, i.e., $L/R$,
$\epsilon_R/\epsilon_L$, $\gamma/(R\epsilon_0)$, $D/R$, and $\lambda_D/R$ such that the resulting
total pair potential $U(\vec{r}_{12},\vec\omega_1,\vec\omega_2)/\epsilon_0$
(compare Eq.~(\ref{eq:Pairpot}) and Fig.~\ref{fig:pairpot}) resembles (qualitatively) the actual pair
potential of the considered ILC molecules. In \MODIFIEDtwo{this} regard, it is worth mentioning that
\MODIFIEDtwo{in principle} comparisons of our effective \MODIFIEDtwo{theory with} particle simulations
can be made, \MODIFIEDtwo{related} to the \MODIFIEDtwo{study} by Saielli et al.~\cite{Saielli2017},
who performed molecular dynamics (MD) simulations \MODIFIEDtwo{for} a mixture of (ellipsoidal) Gay-Berne
and (spherical) Lennard-Jones particles.
\MODIFIEDtwo{Additionally,} both species carry charges and therefore resemble cations
\MODIFIEDtwo{and anions, respectively}.
Our ad-hoc pair potential (Eq.~(\ref{eq:Pairpot})) of the coions can be compared \MODIFIEDtwo{with} the
effective interaction, which can be \MODIFIEDtwo{determined} as the logarithm of the particle-particle
distribution function of the elongated cations in the MD simulations.}

\MODIFIED{We note, that the choices $L/R=4$ and $\epsilon_R/\epsilon_L=2$, which are used throughout
\MODIFIEDtwo{our analysis}, are comparable to \MODIFIEDtwo{those used in} previous studies
\MODIFIEDtwo{(see, e.g., Refs.~\cite{Berardi1993,Kondrat_et_al2010,Bartsch2017,Saielli2017}) for}
similar kinds of particles.
While these values of the Gay-Berne parameters give rise to the formation of smectic phases,
the occurrence of nematic phases is typically observed for much larger values of $L/R$ and
$\epsilon_R/\epsilon_L$~\cite{Bates1999}.}

\subsection{\label{sec:theory:DFT}Density functional theory}
\begin{figure}[!t]
 \includegraphics[width=0.45\textwidth]{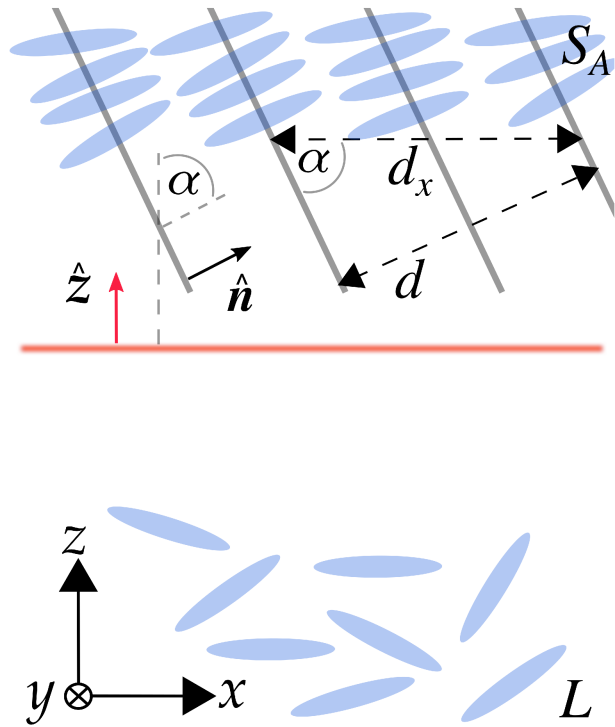}
  \caption{
  Sketch of the \modifiedGreen{interface structure} under consideration.
  Consider a planar interface, illustrated by the horizontal red line, between the isotropic
  \modifiedRed{bulk} liquid $L$, \modifiedRed{imposed} as the boundary condition at $z\rightarrow-\infty$,
  and the smectic-A phase $S_A$ (or $S_{AW}$), \modifiedRed{imposed as} the boundary condition at
  $z\rightarrow+\infty$. Thus, the interface normal
  (red vertical arrow) points \modifiedRed{into the} $z$-direction. At the top, the tails of four
  layers of particles \modifiedRed{of the (ordinary) $S_A$ phase are visible}, which are well aligned
  with the smectic layer normal $\vec{\hat n}:=\sin(\alpha)\,\vec{\hat x}+\cos(\alpha)\,\vec{\hat z}$.
  In the bulk $S_A$ phase, the system is periodic in the direction of the smectic layer normal
  $\vec{\hat n}$ with periodicity $d$ which is a multiple of the smectic layer spacing.
  \modifiedBlue{For the $S_A$ phase $d$ turns out to be two times the distance between neighboring 
  smectic layers \modifiedGreen{(see Sec.~\ref{sec:theory:DFT} below Eq.~(\ref{eq:layernormal}))}}.
  Thus, for a given tilt angle $\alpha$ between the interface normal and the smectic layer normal
  $\vec{\hat n}$, the system is periodic in $x$-direction with periodicity $d_x=d/\sin(\alpha)$.
  Note that the interface \modifiedGreen{structure} is \modifiedRed{translationally invariant in the}
  $y$-direction for all angles $0\leq\alpha\leq\pi/2$. For $\alpha=0$ the system
  \modifiedRed{exhibits in addition} translational invariance in \modifiedRed{the} $x$-direction.
  }
 \label{fig:interface_sketch}
\end{figure}
The degrees of freedom of the particles (compare Sec.~\ref{sec:theory:model}) are fully described by the
positions $\vec{r}$ of their centers and the orientations $\vec\omega$ of their long axes.
Thus, within density functional theory, an appropriate variational grand potential functional
$\beta\Omega[\rho]$ of position- and orientation-dependent number density profiles $\rho(\vec{r},\vec\omega)$
has to be found; \modifiedRed{the equilibrium density profile minimizes the functional.}
The grand potential functional for uniaxial particles,
in the absence of external fields, can generically be expressed as
\begin{equation}
 \begin{split}
    \beta\Omega\left[\rho\right]=
  & \Int{\mathcal{V}}{3}{r}\Int{\mathcal{S}}{2}{\omega}
    \rho(\vec{r},\vec\omega)
    \left[\ln\left(4\pi\Lambda^3\rho(\vec{r},\vec\omega)\right)\right.\\
  &-\left.\left(1+\beta\mu\right)\right]
   +\beta\mathcal{F}\left[\rho\right],
 \end{split}
\label{eq:Omega}
\end{equation}
where the integration domains $\mathcal{V}$ and $\mathcal{S}$ denote the system volume
and the full solid angle, respectively.
The first term in Eq.~(\ref{eq:Omega}) is the purely entropic free energy contribution
of non-interacting uniaxial particles,
where $\beta=1/(k_BT)$ denotes the inverse thermal energy,
$\mu$ the chemical potential, and $\Lambda$ the thermal de Broglie wavelength.

The last term is the excess free energy $\beta\mathcal{F}\left[\rho\right]$ in units of $k_BT$,
which incorporates the effects of the inter-particle interactions. 
Minimizing Eq.~(\ref{eq:Omega}) leads to the Euler-Lagrange equation,
which implicitly determines the equilibrium density profile $\rho(\vec{r},\vec\omega)$:
\begin{equation}
  \rho(\vec{r},\vec\omega)=\frac{
  \exp\left[\beta\mu+c^{(1)}\left(\vec{r},\vec\omega,[\rho]\right)\right]
  }{4\pi\Lambda^3},
\label{eq:ELG}
\end{equation}
where
\begin{equation}
  c^{(1)}\left(\vec{r},\vec\omega,[\rho]\right)=
  -\frac{\delta\beta\mathcal{F}[\rho]}{\delta\rho}
\label{eq:Dir1CorrFunc}
\end{equation}
is the one-particle direct correlation function.
It is fully determined by the excess free energy functional $\beta\mathcal{F}[\rho]$.

The excess free energy functional is the characterizing quantity of the underlying many-body problem.
\modifiedRed{However, in general it is} not known exactly so that one has to \modifiedRed{adopt} appropriate
approximations \modifiedRed{of} it. Following the approach of our previous \modifiedRed{study}~\cite{Bartsch2017}
\modifiedRed{concerning} the bulk phase behavior of \modifiedRed{ILCs, in the spirit of Ref.~\cite{Tarazona1985}
a weighted density expression for} $\beta\mathcal{F}[\rho]$ is considered:
\begin{equation}
  \beta\mathcal{F}[\rho]=
  \frac{1}{2}\Int{\mathcal{V}}{3}{r}\Int{\mathcal{S}}{2}{\omega}
  \rho(\vec{r},\vec\omega)
  \beta\psi\left(\vec{r},\vec\omega,[\bar\rho]\right),
\label{eq:F_WDA}
\end{equation}
where $\beta\psi(\vec{r},\vec\omega,[\bar\rho])$ denotes the effective one-particle potential. 
It is a functional of the so-called projected density $\bar\rho(\vec{r},\vec\omega)$:
%
%
%
%
\begin{equation}
 \begin{split}
  & \bar\rho(\vec{r},\vec\omega,[\rho]) = \frac{1}{4\pi}\bigg[
    Q_0(\vec{r},[\rho])+
    Q_1(\vec{r},[\rho])\cos\left(2\pi(\vec{r}\cdot\vec{\hat n})/d\right)\\
  &+Q_2(\vec{r},[\rho])\cos\left(4\pi(\vec{r}\cdot\vec{\hat n})/d\right)+5P_2(\vec{\omega}\cdot\vec{\hat n})
    \bigg(Q_3(\vec{r},[\rho])\\
  &+Q_4(\vec{r},[\rho])\cos\left(2\pi(\vec{r}\cdot\vec{\hat n})/d\right)
   +Q_5(\vec{r},[\rho])\cos\left(4\pi(\vec{r}\cdot\vec{\hat n})/d\right)\bigg)\bigg],
 \end{split}
\label{eq:WeightedDensity}
\end{equation}
where $P_2(y)=(3y^2-1)/2$ is the Legendre polynomial of degree $2$.
\modifiedRed{We point out} that $\bar\rho(\vec{r},\vec\omega)$ represents an expansion of the density
profile $\rho(\vec{r},\vec\omega)$ in terms of a second-order \modifiedRed{Fourier, and a second-order
Legendre series, respectively}.
Thus, the coefficients $Q_i(\vec{r})$ are the corresponding expansion coefficients,
which will be defined below.
\MODIFIED{It is worth mentioning, that although the projected density $\bar\rho(\vec{r},\vec\omega)$
might take negative values, this does not imply an unphysical behavior as the actual density
$\rho(\vec{r},\vec\omega)$ is \MODIFIEDtwo{determined from} the Euler-Lagrange equation
(i.e., Eq.~(8)) and thus is strictly positive.}
The following three types of bulk phases can be studied within
this \modifiedRed{particular} framework~\cite{Bartsch2017}:
First, isotropic liquids with $Q_0=\text{const}_0$ and $Q_i=0$ for $i>0$.
Second, nematic liquids with $Q_i=\text{const}_i$, if $i=0,3$, and $Q_i=0$ otherwise.
Third, smectic-A phases with $Q_i=\text{const}_i$ for $i\in\{0,\cdots,5\}$.
While \modifiedRed{for isotropic and nematic liquids} the system is translationally invariant
in all spatial directions, in \modifiedRed{the} case of smectic-A phases the system is periodic in the
direction of the smectic layer normal $\vec{\hat n}$ with periodicity $d$,
which is a multiple of the smectic layer spacing.
For smectic-A phases the director is parallel to the smectic layer normal $\vec{\hat n}$ and therefore 
\modifiedRed{the occurrence of} rotationally symmetric distributions of \modifiedRed{the} orientations
$\vec\omega$ around $\vec{\hat n}$, incorporated by the dependence
\modifiedRed{on $\vec\omega\cdot\vec{\hat n}$} in Eq.~(\ref{eq:WeightedDensity}), are plausible.
\modifiedRed{We note} that \modifiedGreen{odd} Fourier-modes in the projected density
$\bar\rho(\vec{r},\vec\omega)$ vanish for bulk smectic-A phases, if the coordinate system is chosen such
that the origin is located \modifiedRed{at} the center of one of the smectic layers due to the mirror
symmetry of smectic layers around their center.
This is a direct consequence of the underlying point symmetry of the particles
considered here \modifiedRed{(see Fig.~\ref{fig:ellipsoids})}. Considering additional terms,
corresponding to the \modifiedGreen{odd} modes in the second-order Fourier expansion of the density
$\rho(\vec{r},\vec\omega)$, would only give rise to a shift of the location of the bulk smectic layers.
Although for systems \modifiedRed{with interfaces} the \modifiedGreen{odd} modes in general do not vanish,
here we neglect these contributions completely. The implications of additionally considering
the \modifiedGreen{odd} terms (up to second order) \modifiedRed{are} discussed in Appendix~\ref{sec:appendix:UnevenModes}. 
Both approaches are weighted-density-like approximations \modifiedRed{of} the exact free energy functional.
A priori, it is not obvious which one leads to better results, because considering more terms of
the Fourier series leads only to a more \modifiedRed{accurate} representation of $\rho(\vec{r},\vec\omega)$
by the projected density $\bar\rho(\vec{r},\vec\omega)$.
However, this does not imply that the resulting free energy functional $\beta\mathcal{F}[\bar\rho]$
is closer to its exact \modifiedRed{form}, because independent of the choice for
$\bar\rho(\vec{r},\vec\omega)$ it relies on the Parsons-Lee approach for its reference part and on the
so-called modified mean-field approximation for the excess part \modifiedRed{(see below)}.
Nevertheless, our approach of considering \modifiedRed{in Eq.~(\ref{eq:WeightedDensity})} only the even
modes up to \modifiedGreen{second order} captures the three types of bulk phases $L$, $N$,
\modifiedRed{as well as} $S_A,S_{AW}$, which are relevant for \modifiedRed{the present} study in the
same way as the full second-order Fourier expansion.

The effective one-particle potential $\beta\psi(\vec{r},\vec\omega)$
consists of two contributions. The first one incorporates the hard-core interactions via
the well-studied Parsons-Lee functional~\cite{Parsons1979,Lee1987},
\begin{align}
  &\beta\psi_\text{PL}(\vec{r},\vec\omega,[\bar\rho]) = 
   -\Int{\mathcal{V}}{3}{r'}\Int{\mathcal{S}}{2}{\omega'}
   \bar\rho(\vec{r}',\vec\omega')\nonumber\\
  &\times\frac{\mathcal{J}(Q_0(\vec{r}))+\mathcal{J}(Q_0(\vec{r}'))}{2}
   f_M(\vec{r}-\vec{r}',\vec\omega,\vec\omega'),
\label{eq:Eff1Pot_PL}
\end{align}
where $f_M(\vec{r}-\vec{r}',\vec\omega,\vec\omega')$ is the Mayer f-function~\cite{Hansen1976}
of the hard core pair interaction potential and $\mathcal{J}(Q_0)$ modifies the corresponding
original Onsager free energy functional (i.e., the second-order virial approximation) such that
the Carnahan-Starling equation of state~\cite{Lee1987} is reproduced for spheres, i.e.,
\modifiedRed{for} $L=R$~\cite{Onsager1949,VanRoij2005}:
\begin{equation}
 \mathcal{J}(Q_0)=\frac{1-\frac{3}{4}\eta_0(Q_0)}{(1-\eta_0(Q_0))^2},
\label{eq:ScalingFunctionJ}
\end{equation}
where \modifiedRed{$\eta_0(Q_0)=Q_0\,LR^2\pi/6$ (for $Q_0$ see Eq.~(\ref{eq:WeightedDensity}))} 
denotes the mean packing fraction within one (bulk) smectic layer.
It is proportional to the coefficient $Q_0$, which gives the mean density within such a smectic layer
(see below). The original Onsager functional is recovered \modifiedRed{in Eq.~(\ref{eq:Eff1Pot_PL})} by
replacing $\mathcal{J}(Q_0)$ by $Q_0$.

The second contribution to the effective one-particle potential $\beta\psi[\bar\rho]$
takes into account the interactions beyond the contact distance
(see the case $|\vec{r}_{12}|\geq R\sigma$ in Eq.~(\ref{eq:Pairpot}))
within the modified mean-field approximation~\cite{Teixeira1991}, \modifiedRed{which is}
a variant of the extended random phase approximation (ERPA)~\cite{Evans1979}:
\begin{align}
  &\beta\psi_\text{ERPA}(\vec{r},\vec\omega,[\bar\rho]) =
   \Int{\mathcal{V}}{3}{r'}\Int{\mathcal{S}}{2}{\omega'}
   \bar\rho(\vec{r}',\vec\omega')\nonumber\\
  &\times\beta U(\vec{r}-\vec{r}',\vec\omega,\vec\omega')
     (1+f_M(\vec{r}-\vec{r}',\vec\omega,\vec\omega')).
\label{eq:Eff1Pot_ERPA}
\end{align}

The present \modifiedRed{study} is devoted to the \modifiedRed{analysis} of free interfaces which are
formed between coexisting bulk phases. In particular, \modifiedRed{the} planar interfaces between the
isotropic liquid $L$ and \modifiedRed{the} two different types of smectic-A phases ($S_A$ or $S_{AW}$,
see Sec.~\ref{sec:results}) will be considered, \modifiedRed{for which} the interface normal 
\modifiedRed{is expected to} be parallel to the $z$-direction (see Fig.~\ref{fig:interface_sketch}).
Due to the isotropy of the liquid phase $L$, the direction of the smectic layer normal
\modifiedRed{
\begin{equation}
\vec{\hat n}(\alpha):=\sin(\alpha)\,\vec{\hat x}+\cos(\alpha)\,\vec{\hat z}
\label{eq:layernormal}
\end{equation}
}
can be chosen to lay in the $x$-$z$-plane. Its orientation is \modifiedRed{fully} determined by the tilt
angle $\alpha$. For $\alpha=0$ the smectic layer normal $\vec{\hat n}=\vec{\hat z}$ points
\modifiedRed{into the} $z$-direction, like the interface normal, while for $\alpha=\pi/2$ it points 
\modifiedRed{into} the $x$-direction and \modifiedRed{thus} it is perpendicular to the interface normal.
The \modifiedRed{interfacial} systems considered here are \modifiedRed{translationally} invariant in the
$y$-direction and show a periodic structure in the $x$-direction with a periodicity $d_x=d/\sin(\alpha)$
\modifiedRed{(compare Fig.~\ref{fig:interface_sketch})} where $d$ is a multiple of the smectic layer spacing.
\modifiedBlue{
(We note that the value of $d$ is determined by the corresponding bulk density \modifiedGreen{distribution}
which minimizes the grand potential functional, i.e., maximizes the bulk pressure
(\modifiedGreen{see} Sec.~2.2.2 in Ref.~\cite{Bartsch2017}). It turns out that, for the $S_{AW}$ phase $d$
equals the smectic layer spacing, while for the $S_A$ phase it equals two times the layer spacing, because
for the $S_A$ phase one obtains bulk solutions $\rho^{(0)}(\vec{r},\vec\omega)$ with $Q_1=Q_4=0$, 
\modifiedGreen{(see, cf., Eqs.~(\ref{eq:WeightedDensity}), (\ref{eq:ExpansionCoeffs_interface}), and
(\ref{eq:ExpansionCoeffs2_interface}))}. Thus the periodicity $d$ along the layer normal $\vec{\hat n}$
is \modifiedGreen{twice} the smectic layer spacing, i.e., the distance between neighboring layers.)
}
\modifiedRed{For} $\alpha=0$, $d_x$ diverges and the system is \modifiedRed{translationally} invariant in
the $x$-direction, too.

As mentioned above, the coefficients $Q_i(\vec{r})$ in Eq.~(\ref{eq:WeightedDensity})
arise from expanding $\rho(\vec{r},\vec\omega)$ in a second-order Legendre-
and Fourier-series~\cite{Bartsch2017}:
\begin{align}
 Q_i(\vec{r},[\rho])&=\frac{1}{\mathcal{V}_d}
 \Int{\mathcal{V}}{3}{r'}\Int{\mathcal{S}}{2}{\omega'}
 \rho(\vec{r}',\vec\omega')w_i(\vec{r},\vec{r}',\vec\omega')
 \label{eq:ExpansionCoeffs_interface}
\end{align}
with
\begin{align}
 w_0&= \mathcal{T}(\vec{r}-\vec{r}'),\nonumber\\
 w_1&=2\mathcal{T}(\vec{r}-\vec{r}')\cos\left(2\pi(\vec{r}'\cdot\vec{\hat n})/d\right),\nonumber\\
 w_2&=2\mathcal{T}(\vec{r}-\vec{r}')\cos\left(4\pi(\vec{r}'\cdot\vec{\hat n})/d\right),\nonumber\\
 w_3&= \mathcal{T}(\vec{r}-\vec{r}')P_2(\vec\omega'\cdot\vec{\hat n}),\nonumber\\
 w_4&=2\mathcal{T}(\vec{r}-\vec{r}')P_2(\vec\omega'\cdot\vec{\hat n})\cos\left(2\pi(\vec{r}'\cdot\vec{\hat n})/d\right),\nonumber\\
 w_5&=2\mathcal{T}(\vec{r}-\vec{r}')P_2(\vec\omega'\cdot\vec{\hat n})\cos\left(4\pi(\vec{r}'\cdot\vec{\hat n})/d\right),
 \label{eq:ExpansionCoeffs2_interface}
\end{align}
where
\begin{equation}
   \mathcal{T}(\vec{r}-\vec{r}')= 
   \begin{cases}
      1,&\vec{r}-\vec{r}'\in\mathcal{V}_d\\
      0,&\text{else}.
   \end{cases}
\label{eq:cut-off-function}
\end{equation}
$\mathcal{T}(\vec{r}-\vec{r}')$ is a cut-off function which defines the integration domain
$\mathcal{V}_d:=\Int{\mathcal{V}}{3}{r'}\mathcal{T}(\vec{r}-\vec{r}')$ around position $\vec{r}$.
For $0<\alpha\leq\pi/2$ the considered interfaces between \modifiedRed{the} isotropic liquid $L$ and
\modifiedRed{the} smectic-A phases $S_A$ or $S_{AW}$ \modifiedRed{exhibit} periodic structures in the
$x$-direction with periodicity $d_x=d/\sin(\alpha)$. Here, $\mathcal{V}_d$ \modifiedRed{is a} slice of
length $d_x$ in $x$-direction \modifiedRed{with a} vanishing extension in $z$-direction 
\modifiedRed{centered} at position $\vec{r}$, i.e.,
$\mathcal{T}(\vec{r}-\vec{r}')=\Theta(d_x/2-|x-x'|)\delta(z-z')$ where $\Theta(x)$ and $\delta(x)$ are the
Heaviside step function and the Dirac delta function, respectively.
\modifiedBlue{
The index $d$ of the integration domain $\mathcal{V}_d$ indicates that $\mathcal{V}_d$ corresponds to
a region which is specified by the periodicity $d$.
}
Due to the translational invariance in
$y$-direction the extension of the integration domain $\mathcal{V}_d$ can be chosen arbitrarily in the
$y$-direction. \modifiedRed{Due to} the periodicity of $\rho(\vec{r},\vec\omega)$ in the $x$-direction, 
this choice of the integration domain $\mathcal{V}_d$ leads to coefficients $Q_i(z)$
(Eq.~(\ref{eq:ExpansionCoeffs_interface})) which depend only on $z$, i.e., on the coordinate parallel
to the interface normal.

For $0<\alpha\leq\pi/2$ one could also consider an integration domain which has a non-vanishing extent
in $z$-direction. \modifiedRed{However,} such a choice has at least two disadvantages:
First, unlike $d_x$, which corresponds to the \modifiedRed{periodicity of the system in $x$-direction,
for $0<\alpha\leq\pi/2$ there is no} obvious choice \modifiedMagenta{for the extent of $\mathcal{V}_d$}
parallel to the interface normal.
Additionally, there is no unique choice for the geometrical \modifiedRed{shape} of the integration domains;
besides using a rectangular form, one could also use any other (\modifiedRed{two}-dimensional) geometrical
object as integration domain $\mathcal{V}_d$.
In this sense the slice of length $d_x$ perpendicular to the interface normal is a simple
but also consistent choice.
Second, this choice \modifiedRed{renders} the evaluation numerically less demanding,
because it requires only a one-dimensional integration
(\modifiedRed{exploiting} the translational invariance in $y$-direction),
instead of evaluating a two-dimensional integral.
\modifiedRed{We note}, that an infinite extent of the integration domain parallel to the interface normal
leads to coefficients $Q_i$ \modifiedRed{which} are independent of the position $\vec{r}$ and therefore
cannot be used to obtain interface profiles.

If $\alpha=0$, i.e., the smectic layer normal $\vec{\hat n}=\vec{\hat z}$ is parallel to the interface
normal, $d_x$ diverges and the the system is \modifiedRed{translationally} invariant in $x$- and
$y$-direction. \modifiedRed{In this case}, the integration domain $\mathcal{V}_d$ \modifiedRed{has an
extent of} length $d$ in $z$-direction, i.e., $\mathcal{T}(\vec{r}-\vec{r}')=\Theta(d/2-|z-z'|)$ with
arbitrary \modifiedRed{extent} in \modifiedRed{the} lateral dimensions $x$ and $y$.
\modifiedRed{As before}, the coefficients $Q_i(z)$ depend only on the $z$-coordinate.
It is worth mentioning, that for all tilt angles $0\leq\alpha\leq\pi/2$ the correct (constant) bulk values
of the coefficients $Q_i$ are \modifiedRed{recovered}, although for $0<\alpha<\pi/2$ the orientation of the
integration domain $\mathcal{V}_d$ (recall that $\mathcal{V}_d$ is a slice of \modifiedRed{width} $d_x$ in
$x$-direction for all $\alpha\in(0,\pi/2]$) changes with respect to the direction of the smectic layer normal
$\vec{\hat n}(\alpha)$. However, because the integration domain \modifiedRed{covers} a full period $d_x$ in
$x$-direction, it gives the same values for the coefficients $Q_i$ in the bulk phases, as for evaluating the
\modifiedRed{coefficients} $Q_i$ with an integration domain parallel to the smectic layer normal $\vec{\hat n}$,
which is the case for $\alpha=0$ and $\pi/2$.

Finally, the one-particle direct correlation function $c^{(1)}\left(\vec{r},\vec\omega,[\rho]\right)$
can be derived by considering Eq.~(\ref{eq:Dir1CorrFunc}) \modifiedRed{which} leads to the following
(modified) expression for $c^{(1)}\left(\vec{r},\vec\omega,[\rho]\right)$
(compare Eq.~(21) in Ref.~\cite{Bartsch2017}): 
\begin{equation}
 \begin{split}
  &c^{(1)}\left(\vec{r},\vec\omega,[\rho]\right)=
   -\beta\psi(\vec{r},\vec\omega,[\bar\rho])+\\
  &\frac{1}{2\mathcal{V}_d}\Int{\mathcal{V}}{3}{r'}\Int{\mathcal{S}}{2}{\omega'}
   \bar\rho(\vec{r}',\vec\omega')\partial_{Q_0}\mathcal{J}(Q_0(\vec{r'}))
   \mathcal{T}(\vec{r}-\vec{r}')\times\\
  &\Int{\mathcal{V}}{3}{r''}\Int{\mathcal{S}}{2}{\omega''}
   \bar\rho(\vec{r}'',\vec\omega'')f_M(\vec{r}'-\vec{r}'',\vec\omega',\vec\omega'').
 \end{split}
\label{eq:Calc_c1_Approx_interface}
\end{equation}
\MODIFIED{
\MODIFIEDtwo{We note} that in Eq.~(\ref{eq:Calc_c1_Approx_interface})
$\frac{\delta Q_0(\vec{r}')}{\delta\bar\rho(\vec{r},\vec\omega)}$ has been replaced by
$\frac{\delta Q_0(\vec{r}')}{\delta\rho(\vec{r},\vec\omega)}=
 \frac{\mathcal{T}(\vec{r}-\vec{r}')}{\mathcal{V}_d}$.
This replacement, i.e., \MODIFIEDtwo{the equation}
$\frac{\delta Q_0(\vec{r}')}{\delta\bar\rho(\vec{r},\vec\omega)}=
 \frac{\delta Q_0(\vec{r}')}{\delta\rho(\vec{r},\vec\omega)}$, is \MODIFIEDtwo{valid exactly} only
for bulk phases.
In general, these two functional derivatives are related via  
$\frac{\delta Q_0(\vec{r}')}{\delta\bar\rho(\vec{r},\vec\omega)}=
\Int{\mathcal{V}}{3}{r''}\Int{\mathcal{S}}{2}{\omega''}
\frac{\delta Q_0(\vec{r}')}{\delta\rho(\vec{r}'',\vec\omega'')}
\frac{\delta\rho(\vec{r}'',\vec\omega'')}{\delta\bar\rho(\vec{r},\vec\omega)}$, which, however,
cannot be calculated analytically.
\MODIFIEDtwo{Determining $\frac{\delta\rho(\vec{r}'',\,\vec\omega'')}{\delta\bar\rho(\vec{r},\,\vec\omega)}$}
requires the functional derivative of the Euler-Lagrange equation (i.e., Eq.~(\ref{eq:ELG}))
which would \MODIFIEDtwo{in turn} produce terms containing
$\frac{\delta Q_0(\vec{r}')}{\delta\bar\rho(\vec{r},\vec\omega)}$.
Nevertheless, the derivation of Eq.~(\ref{eq:Calc_c1_Approx_interface})
(following from Eq.~(21) in Ref.~\cite{Bartsch2017}) incorporates a modification
of the exact one-particle direct correlation \MODIFIEDtwo{function}
such that the density profile $\rho(\vec{r},\vec\omega)$ is replaced by the
projected density $\bar\rho(\vec{r},\vec\omega)$. In this respect, replacing 
$\frac{\delta Q_0(\vec{r}')}{\delta\bar\rho(\vec{r},\vec\omega)}$ by
$\frac{\delta Q_0(\vec{r}')}{\delta\rho(\vec{r},\vec\omega)}=
 \frac{\mathcal{T}(\vec{r}-\vec{r}')}{\mathcal{V}_d}$ (which follows from
Eqs.~(\ref{eq:ExpansionCoeffs_interface})-(\ref{eq:cut-off-function})) is consistent with our approach,
as it also implies \MODIFIEDtwo{an} exchange of $\rho(\vec{r},\vec\omega)$ and
$\bar\rho(\vec{r},\vec\omega)$. 
\MODIFIEDtwo{Moreover, the exchange renders} the correct bulk limit of the interface profile
$\rho(\vec{r},\vec\omega)$ at the boundaries, i.e., $z\rightarrow\pm\infty$.}

\modifiedRed{Equation~(\ref{eq:ELG}) has been} solved numerically (utilizing a Picard scheme with retardation)
\modifiedRed{by} using Eq.~(\ref{eq:Calc_c1_Approx_interface}) \modifiedRed{as well as} the (constant) bulk
values of the coefficients $Q_{i,L}=Q_i(z\rightarrow-\infty)$ in the isotropic liquid phase $L$
and $Q_{i,S}=Q_i(z\rightarrow\infty)$ in the smectic-A phase ($S_A$ or $S_{AW}$) at coexistence
$(T,\mu)=(T_\text{coex},\mu_\text{coex})$. The structural properties and the orientational order
at the free interface are analyzed in terms of the interface profiles of the packing fraction
\modifiedRed{
\begin{equation}
 \eta(\vec{r})=\frac{\pi}{6}LR^2n(\vec{r})
              =\frac{\pi}{6}LR^2\Int{\mathcal{S}}{2}{\omega}\rho(\vec{r},\vec\omega) 
 \label{eq:numberdensity}
\end{equation}
}
\modifiedRed{with the number density $n(\vec{r}):=\Int{\mathcal{S}}{2}{\omega}\rho(\vec{r},\vec\omega) $,}
and \modifiedGreen{in terms of} the orientational order parameter
\modifiedRed{
\begin{equation}
 S_2(\vec{r}):=\Int{\mathcal{S}}{2}{\omega}f(\vec{r},\vec\omega)P_2(\vec\omega\cdot\vec{\hat n});
 \label{eq:oriorderparameter}
\end{equation}
}
$f(\vec{r},\vec\omega):=\rho(\vec{r},\vec\omega)/n(\vec{r})$
\modifiedRed{describes the orientational distribution.}

\subsection{\label{sec:theory:gibbs_div_surface}Gibbs dividing surface}

The position \modifiedRed{$z_\eta$} of the interface is determined by the density profile $\rho(\vec{r},\vec\omega)$
\modifiedRed{for which we have adopted the notion of the} \textit{Gibbs dividing surface}~\cite{Hansen1976}:
\begin{equation}
 h_\eta(z_\eta):=\int_{-\infty}^{z_\eta}dz'(\eta_0(\vec{r}')-\eta_{0,L})+
                 \int_{z_\eta}^{\infty} dz'(\eta_0(\vec{r}')-\eta_{0,S_A})=0,
\label{eq:gibbs_dividing_surface}
\end{equation}
where $\eta_0(\vec{r})=Q_0(\vec{r})\,LR^2\pi/6$ is the mean packing fraction at position $\vec{r}$.
\modifiedRed{The quantities} $\eta_{0,L}=\eta(z\rightarrow-\infty)$ and 
$\eta_{0,S_A}=\eta(z\rightarrow\infty)$ are the bulk values of $\eta_0(\vec{r})$ in the isotropic liquid
phase $L$ and the smectic-A phase $S_A$ (or $S_{AW}$), respectively. The interface position $z_\eta$ in
Eq.~(\ref{eq:gibbs_dividing_surface}) corresponds to the location of a step-like profile
\modifiedRed{such that the number of particles in excess and in deficit of the bulk values is the same
on both sides of the interface.} Taking the derivative of the left-hand side $h_\eta(z)$ of
Eq.~(\ref{eq:gibbs_dividing_surface}) with respect to $z$ 
leads to $h_\eta'(z=z_\eta):=\eta_{0,S_A}-\eta_{0,L}$
which is a constant.
Therefore $h_\eta(z)=(\eta_{0,S_A}-\eta_{0,L})z+h_\eta(0)$ is a linear function
and one has to evaluate $h_\eta(0)$ only once \modifiedRed{in order} to obtain 
\begin{equation}
z_\eta =-h_\eta (0)/(\eta_{0,S_A}-\eta_{0,L}),
\label{eq:gibbs_dividing_surface_evaluation_eta}
\end{equation}
using Eq.~(\ref{eq:gibbs_dividing_surface}), i.e., $h_\eta(z_\eta)=0$.
While $z_\eta$ can be interpreted as the location of the
\textit{transition in the structure} from the isotropic liquid $L$ to the
smectic-A phase $S_A$ (or $S_{AW}$),
replacing $\eta$ by $S_2$ in Eq.~(\ref{eq:gibbs_dividing_surface}) defines a position
\begin{equation}
z_{S_2}=-h_{S_2}(0)/(S_{20,S_A}-S_{20,L}),
\label{eq:gibbs_dividing_surface_evaluation_s2}
\end{equation}
which corresponds to the \textit{transition in the orientational order}
from one phase to the other.

Note, that instead of using the mean packing fraction $\eta_0$ or
the mean orientational order parameter $S_{20}$ in Eq.~(\ref{eq:gibbs_dividing_surface}),
\modifiedRed{for determining the interface positions,} in principle, one could also use the profiles
$\eta(\vec{r})$ and $S_2(\vec{r})$ directly. However, the disadvantage of this \modifiedRed{latter}
approach is that \modifiedRed{in the smectic-A bulk phase $S_A$ (or $S_{AW}$)} the profiles
$\eta(\vec{r})$ and $S_2(\vec{r})$ are still functions of the position $\vec{r}$ (\modifiedRed{via}
the projection $\vec{r}\cdot\vec{\hat n}$ \modifiedRed{onto} the layer normal $\vec{\hat n}$).
\modifiedRed{Typically, this prevents the use of the latter generalized
Eqs.~(\ref{eq:gibbs_dividing_surface_evaluation_eta}) and (\ref{eq:gibbs_dividing_surface_evaluation_s2})
for determining $z_\eta$ and $z_{S_2}$.} Instead, one \modifiedRed{has to} solve
Eq.~(\ref{eq:gibbs_dividing_surface}) numerically, which requires many iterations depending on the
desired accuracy.

Nevertheless, \modifiedRed{in} the particular case $\alpha=\pi/2$ the interface normal and the smectic
layer normal are perpendicular. \modifiedRed{Due} to the translational invariance of \modifiedRed{the}
smectic phases perpendicular to their layer normal, here the density profile
$\eta(z\rightarrow\infty)$ and the orientational order parameter profile $S_2(z\rightarrow\infty)$ do
not depend on $z$ for $z\rightarrow\infty$ in the smectic bulk, but \modifiedRed{they depend}
only on the $x$-coordinate. Thus, for $\alpha=\pi/2$ one can define interface contours
$\tilde z_\eta(x)$ and $\tilde z_{S_2}(x)$, analogously to $z_\eta$ and $z_{S_2}$:
\begin{align}
 \tilde h_m(\tilde z_m(x)):=&
 \int_{-\infty}^{\tilde z_m(x)}dz'(m(\vec{r}')-m_{L})+\nonumber\\&
 \int_{z_m(x)}^{\infty} dz'(m(\vec{r}')-m_{S_A})=0,\nonumber\\
 \tilde z_m(x)=&-\tilde h_m(0)/(m_{S_A}(x)-m_{L}),
\label{eq:gibbs_dividing_surface_evaluation_contours}
\end{align}
where $m\in\{\eta,S_2\}$.

\subsection{\label{sec:theory:interface_tension}Interfacial tension}

The interfacial tension $\Gamma$ is a measure of the \modifiedRed{excess} amount of work needed to
form an interface between coexisting bulk phases~\cite{Hansen1976}. \modifiedRed{Accordingly}, it can
be calculated by \modifiedRed{determining} the increase in the grand potential $\beta\Omega[\rho]$ of
the interface system \modifiedRed{in excess of} the bulk grand potential
$\beta\Omega_0:=-\beta p\mathcal{V}$ which is given by the bulk pressure $p$ (see Eq.~(26) in
Ref.~\cite{Bartsch2017}) times the system volume $\mathcal{V}$:
\begin{equation}
\Gamma^*(\alpha):=\beta\Gamma(\alpha)=
\frac{\beta\Omega([\rho],\alpha)+\beta p_\text{coex}\mathcal{V}}{A},
\label{eq:interface_tension}
\end{equation}
where $A$ is the cross-sectional area of the system in lateral directions to the interface normal.
\modifiedBlue{Hence, $\Gamma^*(\alpha)$ has the dimension $1/\text{area}$.}
The pressure $p_\text{coex}:=p(T_\text{coex},\mu_\text{coex},d)$ at coexistence
$(T,\mu)=(T_\text{coex},\mu_\text{coex})$ is the same in the isotropic liquid $L$ and the smectic-A
phase $S_A$ or $S_{AW}$ with the equilibrium layer spacing $d$.
\modifiedRed{The equilibrium tilt angle $\alpha_\text{eq}$ minimizes} the interfacial tension
$\Gamma^*(\alpha=\alpha_\text{eq})$ (see Sec.~\ref{sec:results:tilt}).

\section{\label{sec:results}Results}

In this section \modifiedRed{we present} results for free interfaces formed between the isotropic
liquid $L$ and \modifiedRed{the} smectic-A phase $S_A$ or $S_{AW}$.
The discussion focuses on two kinds of ionic liquid crystals (ILCs) which
are described by the pair interaction potential $U(\vec{r}_{12},\vec\omega_1,\vec\omega_2)$
\modifiedRed{(Eq.~(\ref{eq:Pairpot}))}, introduced in Sec.~\ref{sec:theory:model}:
First, ILCs with charges in the center, i.e., $D=0$ (see Fig.~\ref{fig:ellipsoids} and
Eq.~(\ref{eq:PairPot_ES})), and second, ILCs with charges at the tips, i.e., $D/R=1.8$.
In particular the structural and orientational properties of the interface are discussed in terms of 
the packing fraction profile $\eta(\vec{r})$ and the orientational order parameter profile
$S_2(\vec{r})$ for \modifiedRed{various} relative orientations between the interface normal and
the smectic layer normal, i.e., for different tilt angles $\alpha$
(see Fig.~\ref{fig:interface_sketch}). All results presented \modifiedRed{here have been} obtained via
the density functional approach described in Sec.~\ref{sec:theory:DFT}.

\subsection{\label{sec:results:para}Interface normal parallel to the smectic layer normal ($\alpha=0$)}
\begin{figure}[!t]
 \includegraphics[width=0.45\textwidth]{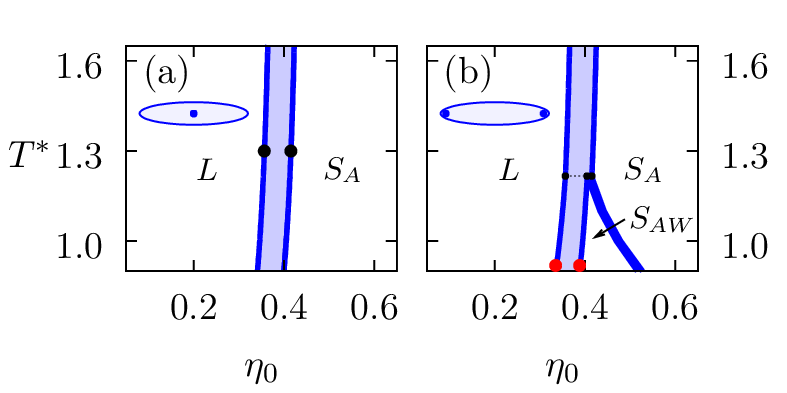}
  \caption{
  Bulk phase diagrams for (a) ionic liquid crystals with
  $L/R=4,\epsilon_R/\epsilon_L=2,\gamma/(R\epsilon_0)=0.045,\lambda_D/R=5$, and $D=0$
  and (b) with
  $L/R=4,\epsilon_R/\epsilon_L=2,\gamma/(R\epsilon_0)=0.045,\lambda_D/R=5$, and $D/R=1.8$.
  For $D=0$, \modifiedRed{i.e., the charges \modifiedGreen{being} concentrated in the center of the molecules,}
  solely a first-order phase transition from the isotropic liquid phase $L$ to the ordinary smectic-A
  phase $S_A$ occurs at sufficiently high mean packing fractions $\eta_0$.
  The ordinary smectic-A phase $S_A$ is characterized by a layer structure with smectic layer spacing
  $d/R\approx4.3\gtrsim L/R=4$ comparable \modifiedRed{with} the particle length $L$. The particles
  in the layers are well aligned with the layer normal $\vec{\hat n}$.
  In panel (b), i.e., for $D/R=1.8$ \modifiedRed{(the charges \modifiedGreen{being} located at the tips
  of the molecules)}, another smectic-A structure, referred to as \modifiedRed{the} $S_{AW}$ phase can
  be observed at low reduced temperatures $T^*$.
  The $S_{AW}$ phase \modifiedRed{exhibits} an alternating structure,
  \modifiedRed{consisting of primary} layers of particles being parallel to the layer normal
  and secondary layers \modifiedRed{in which} the particles prefer to be perpendicular to it.
  This leads to an increased layer spacing $d/R\geq7.5$. The black dotted line in panel (b) marks
  the triple point at $T^*\approx1.23$ \modifiedRed{for} which the isotropic liquid $L$, the ordinary
  smectic-A \modifiedRed{phase} $S_A$, and the $S_{AW}$ phase are in three-phase-coexistence.
  A detailed description of the structural properties of the smectic-A phases $S_A$ and $S_{AW}$,
  including illustrations of their microstructure, are provided in Ref.~\cite{Bartsch2017}.
  \modifiedRed{The black dots ($\bullet$) in panel (a)},
  \modifiedRed{respectively the red dots (\textcolor{red}{$\bullet$}) in (b), mark the coexisting
  bulk states at the reduced temperature $T^*=1.3$, respectively $0.9$, imposed as boundary conditions
  for the free interfaces shown in Figs.~\ref{fig:if_ilc_l-sa_para} and
  \ref{fig:if_ilc_l-saw_para}-\ref{fig:if_ilc_perp_SAW}.}
  }
 \label{fig:pd_ilc}
\end{figure}
\begin{figure}[!t]
 \includegraphics[width=0.45\textwidth]{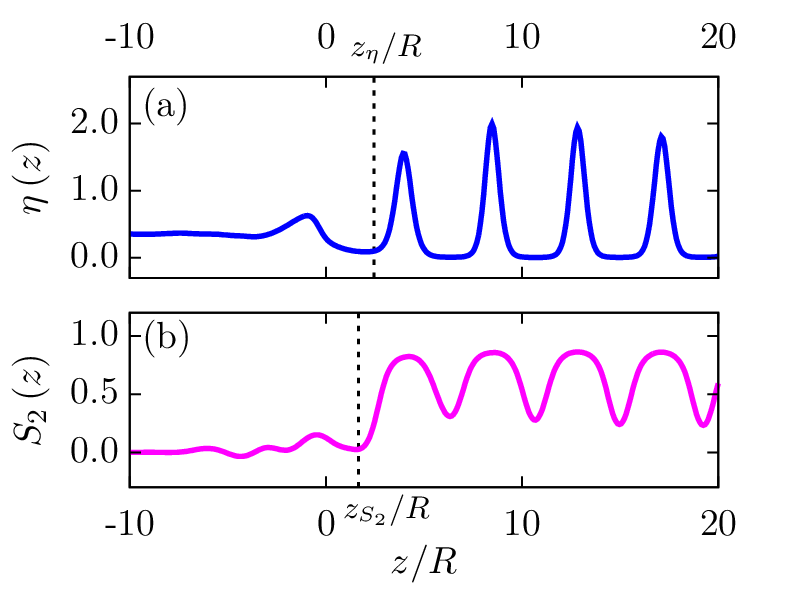}
  \caption{
  The $L$-$S_A$-interface profile of the packing fraction $\eta(z)$, panel (a), and the 
  orientational order parameter $S_2(z)$, panel (b), are shown for an ionic liquid crystal with
  $L/R=4,\epsilon_R/\epsilon_L=2,\gamma/(R\epsilon_0)=0.045,\lambda_D/R=5$, and $D=0$, i.e.,
  \modifiedRed{the} charges \modifiedRed{are concentrated} in the center of the molecules.
  The free interface between the isotropic liquid $L$ (\modifiedRed{imposed} as boundary condition
  for $z\rightarrow-\infty$) and the ordinary smectic-A phase $S_A$
  (\modifiedRed{i.e.,} $z\rightarrow\infty$) is considered for \modifiedRed{the} reduced temperature
  $T^*=1.3$. The corresponding coexisting bulk states are marked by the black dots ($\bullet$)
  in the phase diagram in Fig.~\ref{fig:pd_ilc}(a). \modifiedRed{The} tilt angle \modifiedRed{is}
  $\alpha=0$, i.e., the smectic layer normal $\vec{\hat n}=\vec{\hat z}$ is parallel
  to the interface normal (see Fig.~\ref{fig:interface_sketch}). For $z/R>0$ the last layers of the
  $S_A$ phase \modifiedRed{are visible, in which} the particles are still well aligned with the
  $z$-axis, indicated by large values of the orientational order parameter $S_2(z/R)>0.8$ within
  these layers. For $z/R<0$ the layer structure \modifiedRed{of the density dies out} rapidly and the
  orientational order vanishes as well. Ultimately, the isotropic bulk limit will be approached for
  $z\rightarrow-\infty$. However, already for $z/R<-10$ the profiles \modifiedGreen{have de facto reached}
  their bulk limits in the isotropic liquid $L$. The black dashed lines refer to the interface positions
  $z_\eta$ and $z_{S_2}$, \modifiedRed{respectively,} calculated via Eqs.~(\ref{eq:gibbs_dividing_surface_evaluation_eta})
  and (\ref{eq:gibbs_dividing_surface_evaluation_s2}). The \modifiedRed{difference}
  $(z_\eta-z_{S_2})/R\approx2.45-1.66=0.79$ \modifiedRed{between} the two interface positions is
  considerably smaller than the smectic layer spacing $d/R\approx4.28\gtrsim L/R=4$.
  \modifiedRed{Therefore} the orientational order of the $S_A$ phase vanishes within the last smectic
  layer while approaching the isotropic liquid $L$.
  }
 \label{fig:if_ilc_l-sa_para}
\end{figure}
\begin{figure}[!t]
 \includegraphics[width=0.45\textwidth]{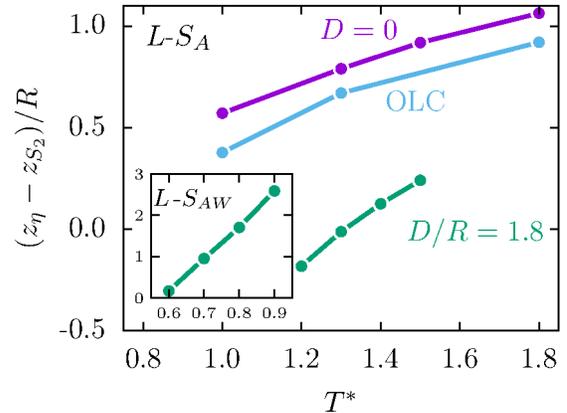}
  \caption{
  The \modifiedRed{difference} $(z_\eta-z_{S_2})/R$ between the Gibbs dividing surface position
  $z_\eta$ \modifiedRed{(Eq.~(\ref{eq:gibbs_dividing_surface_evaluation_eta}))}, and the surface
  position $z_{S_2}$ \modifiedRed{(Eq.~(\ref{eq:gibbs_dividing_surface_evaluation_s2})), which
  corresponds} to the transition \modifiedRed{of} the orientational order at the interface, are
  shown for \modifiedRed{three cases}. First, an ordinary (uncharged) liquid crystal (OLC; blue curve);
  second, \modifiedRed{ILCs} with charges in \modifiedRed{their} center, i.e., $D=0$ (violet curve);
  and, third, \modifiedRed{ILCs} with charges at the tips, i.e., $D/R=1.8$ (green curve).
  Here, the smectic layer normal $\vec{\hat n}=\vec{\hat z}$ is parallel to the interface normal, i.e.,
  $\alpha=0$. In all cases studied, the \modifiedRed{differences} $(z_\eta-z_{S_2})/R$ are smaller than
  the smectic layer spacing $d\gtrsim L$, which \modifiedRed{for the $S_A$ phase} is comparable to the
  particle length $L/R=4$. Thus, the loss of orientational order occurs within the last smectic layer
  before approaching the isotropic liquid $L$. The inset shows data for the $L$-$S_{AW}$-interface,
  \modifiedRed{which are accessible for $D/R=1.8$} at sufficiently low temperatures $T^*$.
  Although the \modifiedRed{difference} $(z_\eta-z_{S_2})/R$ is enlarged for $0.7<T^*\leq0.9$,
  it is still considerably smaller than the layer spacing $d/R\approx7.5$ and decreases
  \modifiedRed{rapidly upon} decreasing \modifiedRed{the} temperature $T^*$. Hence, for $\alpha=0$,
  the orientational order of the smectic-A phase, either $S_A$ or $S_{AW}$, vanishes
  \modifiedRed{directly} with the \modifiedRed{disappearance} of the layer structure at the interface.
  }
 \label{fig:gibbs_parallel}
\end{figure}
\begin{figure}[!t]
 \includegraphics[width=0.45\textwidth]{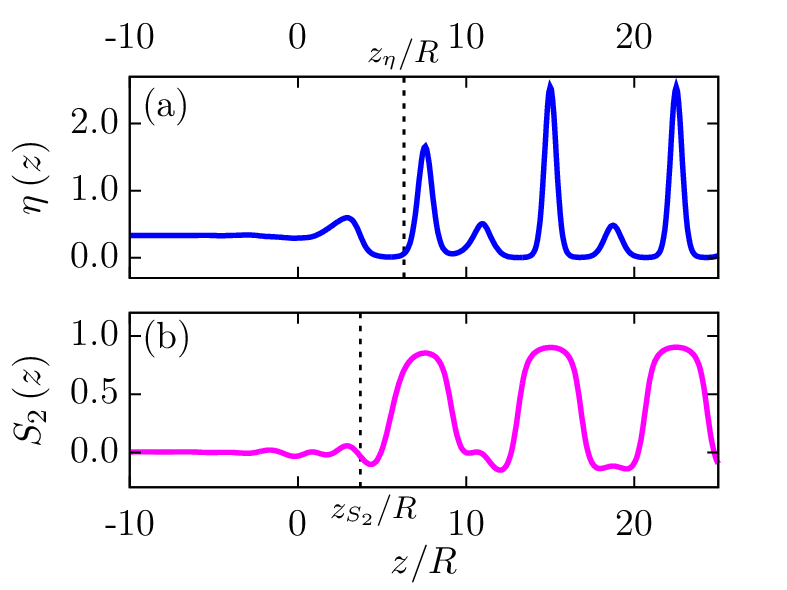}
  \caption{
  \modifiedRed{For $\alpha=0$,} the $L$-$S_{AW}$-interface profiles $\eta(z)$ and $S_2(z)$
  \modifiedRed{are shown} for \modifiedRed{ILCs} with charges at the tips
  ($L/R=4,\epsilon_R/\epsilon_L=2,\gamma/(R\epsilon_0)=0.045,\lambda_D/R=5$, and $D/R=1.8$)
  at \modifiedRed{the} reduced temperature $T^*=0.9$ (see the red dots (\textcolor{red}{$\bullet$})
  in Fig.~\ref{fig:pd_ilc}(b)).
  For $z\rightarrow-\infty$ the isotropic liquid bulk $L$ is approached \modifiedRed{whereas} for
  $z\rightarrow\infty$ the $S_{AW}$ bulk \modifiedRed{is attained}. The \modifiedRed{difference}
  $(z_\eta-z_{S_2})/R\approx6.31-3.72=2.59$ \modifiedRed{between} the two interface positions is
  larger \modifiedRed{than the one of} the $L$-$S_A$-interface
  (compare Figs.~\ref{fig:if_ilc_l-sa_para} and \ref{fig:gibbs_parallel}) but it is still smaller than
  the smectic layer spacing $d/R=7.5$. Therefore the orientational order of the $S_{AW}$ phase also
  vanishes within the range of the last smectic layer at the interface. 
  }
 \label{fig:if_ilc_l-saw_para}
\end{figure}

First, \modifiedRed{we consider the case that} the interface normal \modifiedRed{is} parallel to
\modifiedRed{the normal of the smectic layers}, i.e., $\alpha=0$ (see Fig.~\ref{fig:interface_sketch}).
Both point \modifiedRed{into the} $z$-direction and due to translational invariance in \modifiedRed{the
$x$- and $y$-directions}, the packing fraction $\eta(z)$ and the orientational order parameter $S_2(z)$
are functions solely of the spatial coordinate $z$. For the case of an ionic liquid crystal with
$L/R=4,\epsilon_R/\epsilon_L=2,\gamma/(R\epsilon_0)=0.045,\lambda_D/R=5$, and $D=0$, i.e., the charges
are localized in the center of the molecule, the bulk phase behavior is shown in the
$T^*$-$\eta_0$-phase diagrams of Fig.~\ref{fig:pd_ilc}(a) where $T^*=kT/\epsilon_0$ and
$\eta_0=Q_0\,LR^2\pi/6$ are the reduced temperature and the mean packing fraction, respectively. 
\modifiedRed{Within} the considered temperature range $T^*\in[0.9,1.65]$ solely a first-order phase
transition from the isotropic liquid phase $L$ to the ordinary smectic-A phase $S_A$ occurs.
The $S_A$ phase is characterized by a layer structure with a smectic layer spacing $d/R\approx4.3$,
which is comparable to the particle length $L/R=4$. Within the smectic layers the particles
are well aligned with the smectic layer normal $\vec{\hat n}$. The blue lines in
Fig.~\ref{fig:pd_ilc}(a) correspond to $L$-$S_A$-coexistence and the light blue area in between the
coexistence lines represents the two-phase region.

The $L$-$S_A$-interface is shown in Fig.~\ref{fig:if_ilc_l-sa_para} for $T^*=1.3$.
\modifiedRed{In the phase diagram in Fig.~\ref{fig:pd_ilc}(a)} the corresponding \modifiedRed{two}
coexisting bulk states are marked by black dots ($\bullet$).
\modifiedRed{Panels (a) and (b) show} the packing fraction profile $\eta(z)$ along the interface normal
and the orientational order parameter profile $S_2(x)$\modifiedRed{, respectively}. The black dashed
vertical line in panel (a) marks the position $z_\eta$ of the \modifiedRed{Gibbs} dividing surface,
which is defined by Eq.~(\ref{eq:gibbs_dividing_surface_evaluation_eta}).
Correspondingly, the black dashed vertical line in panel (b)
marks the position $z_{S_2}$ (Eq.~(\ref{eq:gibbs_dividing_surface_evaluation_s2})). 
Apparently, the two interface positions $z_\eta$ and $z_{S_2}$, which are related to the interfacial
transition in the structure and in the orientational order, respectively,
\modifiedRed{differ from each other}. In Fig.~\ref{fig:gibbs_parallel}, these \modifiedRed{differences}
$z_\eta-z_{S_2}$ are plotted as function of \modifiedRed{the} reduced temperature $T^*$ for three
different kinds of liquid-crystalline systems. The violet curve corresponds to \modifiedRed{ILCs}
with \modifiedRed{all charges concentrated} in the molecular centers, i.e., $D=0$, while the green
curve shows data points for $D/R=1.8$. The blue curve corresponds to a system of ordinary (uncharged)
liquid crystals (OLCs) described by $L/R=4$, $\epsilon_R/\epsilon_L=2$, and $\gamma/(R\epsilon_0)=0$.
\modifiedRed{The phase diagram for OLCs is not shown here; it is presented} in Fig.~4(a) of
Ref.~\cite{Bartsch2017}. Within the considered temperature ranges, \modifiedRed{in} all three cases
the \modifiedRed{differences} are at most as large as the length of \modifiedRed{the} particle diameter
$R$, which \modifiedRed{in turn} \modifiedGreen{is} much smaller than the smectic layer spacing
$d/R\approx4.3$ which is comparable to the particle length $L$, because the particles \modifiedRed{within}
the smectic layers are well aligned with the $z$-direction, indicated by $S_2(z)>0.8$ in the centers of
the smectic layers.
Thus, \modifiedRed{the small size of the differences shows that in these cases} the transition in the
orientational order and \modifiedRed{in the} fluid structure go along with each other. As soon as the
smectic layer structure \modifiedRed{dies out}, the orientational order vanishes as well.

While for ILCs with charges in \modifiedRed{their} center, \modifiedRed{within the considered
temperature range}, only $L$-$S_A$-coexistence is observable (see Fig.~\ref{fig:pd_ilc}(a)).
\modifiedGreen{For} ILCs with the charges at the tips, such as \modifiedGreen{in the case}
$L/R=4,\epsilon_R/\epsilon_L=2,\gamma/(R\epsilon_0)=0.045,\lambda_D/R=5$, and $D/R=1.8$,
the bulk phase behavior changes significantly at low temperatures\modifiedGreen{, i.e., for} $T^*<1.23$.
The bulk phase diagram in Fig.~\ref{fig:pd_ilc}(b) shows that \modifiedRed{in this case} the distinct
smectic-A phase $S_{AW}$ occurs for intermediate mean packing fraction $\eta_0$.
The $S_{AW}$ phase is characterized by an alternating layer structure of smectic layers 
with a majority of particles being oriented parallel to the smectic layer normal $\vec{\hat n}$
and a minority of particles localized in secondary layers which prefer orientations
perpendicular to the smectic layer normal.
Due to \modifiedRed{this} alternating layer structure the smectic layer spacing $d/R\approx7.5$
is increased for the $S_{AW}$ phase.
A detailed discussion of the structural and orientational properties of this new and peculiar
smectic-A phase, in particular \modifiedRed{concerning} the bulk density and \modifiedRed{the}
orientational order parameters profiles, is given in Ref.~\cite{Bartsch2017}.

In Fig.~\ref{fig:if_ilc_l-saw_para} the $L$-$S_{AW}$-interface profiles $\eta(z)$  
and $S_2(z)$ are shown for $\alpha=0$ and $T^*=0.9$. \modifiedRed{In the phase diagram in
Fig.~\ref{fig:pd_ilc}(b)} the corresponding coexisting bulk states are marked by red dots
(\textcolor{red}{$\bullet$}). \modifiedRed{On} the right hand side of Fig.~\ref{fig:if_ilc_l-saw_para}
the alternating layer structure of the bulk $S_{AW}$ phase is evident. In the main layers the majority
of \modifiedRed{the} particles ($\eta(z)>2$) has orientations parallel to the $z$-axis ($S_2(z)>0.8$)
and in the secondary layers, \modifiedRed{formed by} less \modifiedRed{of them} ($\eta(z)\approx0.6$),
the particles prefer orientations perpendicular to the $z$-axis ($S_2(z)<0$).
For the $L$-$S_{AW}$-interface the \modifiedRed{difference $(z_\eta-z_{S_2})/R\approx2.6$ of} the two
interface positions is increased compared to the $L$-$S_A$-interface (see Fig.~\ref{fig:gibbs_parallel}),
because the smectic layer spacing $d/R\geq7.5$ in the $S_{AW}$ phase \modifiedRed{is enlarged, too.}
\modifiedRed{As before, the orientational order directly} vanishes with the \modifiedRed{disappearance}
of the layer structure. Furthermore, the inset in Fig.~\ref{fig:gibbs_parallel} shows that
$(z_\eta-z_{S_2})/R$ decreases \modifiedRed{upon} lowering the temperature. \modifiedBlue{Thus the
difference \modifiedGreen{$z_\eta-z_{S_2}$} becomes smaller relative to the layer spacing $d$,
such that the direct vanishing of the orientational order \modifiedGreen{associated} with the
disappearance of the layer structure is observable for the whole temperature range considered here}.

\subsection{\label{sec:results:perp}Interface normal perpendicular to the smectic layer normal ($\alpha=\pi/2$)}
\begin{figure}[!t]
 \includegraphics[width=0.45\textwidth]{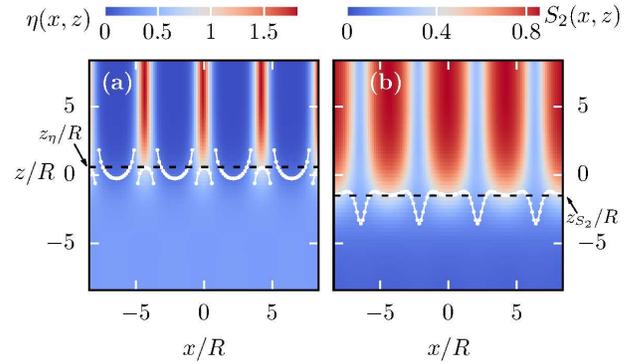}
  \caption{
  The $L$-$S_A$-interface profiles $\eta(x,z)$, panel (a), and $S_2(x,z)$, panel (b),
  are shown for $T^*=1.3$ (see the black dots ($\bullet$) in Fig.~\ref{fig:pd_ilc}(a))
  and $\alpha=\pi/2$. \modifiedRed{Accordingly}, the smectic layer normal $\vec{\hat n}=\vec{\hat x}$
  and the interface normal (parallel to the $z$-axis) are perpendicular.
  Here, \modifiedRed{ILCs} with charges \modifiedRed{at} the center are considered,
  described by the parameter set
  $L/R=4,\epsilon_R/\epsilon_L=2,\gamma/(R\epsilon_0)=0.045,\lambda_D/R=5$, and $D=0$.
  For $z\rightarrow-\infty$ the isotropic \modifiedRed{bulk} liquid $L$ and 
  for $z\rightarrow\infty$ the bulk of the $S_A$ phase is approached.
  The \modifiedRed{decaying} red stripes at the \modifiedRed{upper part of these} plots show the
  tails of the smectic layers located at $x/R\approx0,\pm d/R,\pm2d/R$
  where $d/R\approx4.28$ is the smectic layer spacing.
  The black dashed lines mark the interface positions $z_\eta$ and $z_{S_2}$
  calculated via Eqs.~(\ref{eq:gibbs_dividing_surface_evaluation_eta}) and
  (\ref{eq:gibbs_dividing_surface_evaluation_s2}), while the white dotted lines
  mark the interface contours $\tilde z_\eta(x)$ and $\tilde z_{S_2}(x)$
  calculated via Eq.~(\ref{eq:gibbs_dividing_surface_evaluation_contours}).
  The \modifiedRed{difference} $(z_\eta-z_{S_2})/R\approx0.58-(-1.51)=2.09$ is larger than the particle
  diameter $R$, \modifiedRed{which is} the relevant geometrical property of the particles at this
  interface, because for $\alpha=\pi/2$ the particles in the $S_A$ layers are well aligned with the
  $x$-axis and therefore they are oriented perpendicular to the direction of the interface normal.
  The orientational order of the smectic-A phase persists up to a few particle diameters into the
  liquid phase, unlike the case $\alpha=0$, \modifiedRed{in which the disappearance} of the layer
  structure causes a \modifiedRed{direct} vanishing of the orientational order within the last layer
  (see Figs.~\ref{fig:if_ilc_l-sa_para}-\ref{fig:if_ilc_l-saw_para}).
  }
 \label{fig:if_ilc_perp_SA}
\end{figure}
\begin{figure}[!t]
 \includegraphics[width=0.45\textwidth]{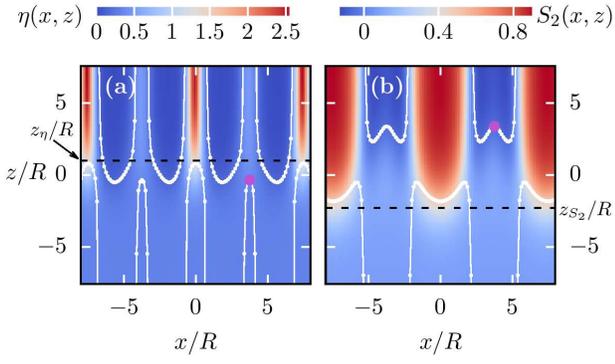}
  \caption{
  The interface profiles $\eta(x,z)$ and $S_2(x,z)$ for $T^*=0.9$ and $\alpha=\pi/2$.
  Here the $L$-$S_{AW}$ interface (see the red dots (\textcolor{red}{$\bullet$})
  in Fig.~\ref{fig:pd_ilc}(b)) for an \modifiedRed{ILC} with \modifiedRed{the} charges at the tips
  ($L/R=4,\epsilon_R/\epsilon_L=2,\gamma/(R\epsilon_0)=0.045,\lambda_D/R=5$, and $D/R=1.8$)
  is considered. The thin red areas in panel (a) for lateral positions $x/R=0,\pm d/R=\pm7.5$
  show the tails of the smectic layers where the particles prefer \modifiedRed{an orientation parallel}
  to the smectic layer normal $\vec{\hat n}=\vec{\hat x}$. \modifiedRed{This is indicated by the} large
  value of $S_2(x,z)>0.8$ within these layers. \modifiedRed{In panel (a)} the secondary layers of the
  $S_{AW}$ phase are shown as light blue areas in panel (a) located at $x/R=\pm d/(2R)=\pm3.75$.
  \modifiedRed{There}, the orientational order parameter $S_2(x,z)$, \modifiedRed{shown in} panel (b),
  is negative. The black dashed lines mark the interface positions $z_\eta$ and $z_{S_2}$
  calculated via Eqs.~(\ref{eq:gibbs_dividing_surface_evaluation_eta})
  and (\ref{eq:gibbs_dividing_surface_evaluation_s2}), while the white dotted lines mark the interface
  contours $\tilde z_\eta(x)$ and $\tilde z_{S_2}(x)$, which have been calculated via
  Eq.~(\ref{eq:gibbs_dividing_surface_evaluation_contours}).
  The \modifiedRed{differences} $(z_\eta-z_{S_2})/R\approx1.0-(-2.3)=3.3$, respectively
  $(\tilde z_\eta(x)-\tilde z_{S_2}(x))/R\approx0.81-(-1.83)=2.64$ \modifiedRed{at the} lateral positions
  $x/R\approx0,\pm7.5$, \modifiedRed{exhibit} a persisting orientational order for the main layers,
  similar to the findings \modifiedRed{for} the $L$-$S_A$ interface
  (compare Fig.~\ref{fig:if_ilc_perp_SA}).  
  Interestingly, at the secondary layers ($x/R=\pm d/(2R)=\pm3.75$) the orientational order
  vanishes \modifiedRed{ahead of the disappearance} of the layer structure, i.e.,
  $\tilde z_{S_2}(x/R=\pm3.75)/R\approx3.39>-0.34\approx\tilde z_{\eta}(x/R=\pm3.75)/R$.
  \modifiedRed{In order to} guide the eye, the magenta dots (\textcolor{Magenta}{$\bullet$}) mark the
  positions $(x/R,\tilde z_{\eta}/R)\approx(3.75,-0.34)$ and $(x/R,\tilde z_{S_2}/R)\approx(3.75,3.39)$.
  }
 \label{fig:if_ilc_perp_SAW}
\end{figure}

For $\alpha=\pi/2$ the interface normal and the smectic layer normal are perpendicular to each other.
The smectic layer normal points \modifiedRed{into} the $x$-direction and the interface normal
\modifiedRed{into} the $z$-direction (see Fig.~\ref{fig:interface_sketch}).
The associated $L$-$S_A$-interface at $T^*=1.3$ for an \modifiedRed{ILC system} with 
\modifiedRed{the charges concentrated at the center}, described by the parameter set
$L/R=4,\epsilon_R/\epsilon_L=2,\gamma/(R\epsilon_0)=0.045,\lambda_D/R=5$, and $D=0$,
is shown in Fig.~\ref{fig:if_ilc_perp_SA}. The corresponding bulk phases are given by the state
points marked by black dots ($\bullet$) in the phase diagram in Fig.~\ref{fig:pd_ilc}(a).
Panel (a) shows the packing fraction $\eta(x,z)$ and (b) the orientational order parameter $S_2(x,z)$.
The red areas at the top of Fig.~\ref{fig:if_ilc_perp_SA}(a) show the tails of four smectic layers of
the $S_A$ phase located at \modifiedRed{$x/R=\pm d/(2R)\approx\pm2.14$ and
$x/R=\pm3d/(2R)\approx\pm6.42$} where $d/R\approx4.28$ is the smectic layer spacing.
The particles are well aligned with the smectic layer normal $\vec{\hat n}=\vec{\hat x}$
indicated by large values of the orientational order parameter $S_2(x,z)>0.8$ in the layers. 

The black dashed lines in Fig.~\ref{fig:if_ilc_perp_SA} show the interface positions $z_\eta$
and $z_{S_2}$ calculated \modifiedRed{from} Eqs.~(\ref{eq:gibbs_dividing_surface_evaluation_eta})
and (\ref{eq:gibbs_dividing_surface_evaluation_s2}), while the white dotted lines show the interface
contours $\tilde z_\eta(x)$ and $\tilde z_{S_2}(x)$ obtained \modifiedRed{from}
Eq.~(\ref{eq:gibbs_dividing_surface_evaluation_contours}). The contour lines
$\tilde z_\eta(x)$ and $\tilde z_{S_2}(x)$ at the centers of the tails of the smectic layers, e.g.,
at $x/R\approx2.14$, are very close to $z_\eta$ and $z_{S_2}$, respectively.
This suggests that the two distinct definitions of the interface positions, i.e.,
using either Eqs.~(\ref{eq:gibbs_dividing_surface_evaluation_eta}) and
(\ref{eq:gibbs_dividing_surface_evaluation_s2}) or
Eq.~(\ref{eq:gibbs_dividing_surface_evaluation_contours}), are consistent with each other,
because the majority of the particles in the smectic phase are located close to the centers
of the smectic layers. \modifiedRed{In Fig.~\ref{fig:if_ilc_perp_SA}(a)} the packing fraction
interface contour $\tilde z_\eta(x)$ \modifiedRed{exhibits} discontinuities for lateral positions
$\check{x}$ at which the smectic bulk packing fraction
$\eta_{S_A}(\check{x}):=\eta(\check{x},z\rightarrow\infty)$ takes the same value
$\eta_L=\eta(\check{x},z\rightarrow-\infty)$ as in the isotropic liquid $L$, i.e.,
$\eta_{S_A}(\check{x})=\eta_L$. Thus, the numerical calculation of the Gibbs dividing surface
via Eq.~(\ref{eq:gibbs_dividing_surface_evaluation_contours}) leads to a divergence due to the
vanishing denominator. This can be considered \modifiedRed{as} an artifact, which, however, occurs
only at the particular lateral positions $\check{x}$.
Nevertheless, the benefit of considering $\tilde z_\eta(x)$ and $\tilde z_{S_2}(x)$
as interface positions is their \modifiedRed{dependence} on the lateral coordinate $x$.
In particular, for the case of the $L$-$S_{AW}$-interface it is necessary to consider
$\tilde z_\eta(x)$ and $\tilde z_{S_2}(x)$ in order to study the interface
at the \modifiedRed{main layers and at} the secondary layers separately (see below).

Interestingly, if the layer normal and the interface normal are perpendicular, one observes a
significant \modifiedRed{difference} $(z_\eta-z_{S_2})/R\approx0.72-(-1.76)=2.48$ \modifiedRed{between}
the interface position $z_\eta$, corresponding to the structural transition, and $z_{S_2}$
corresponding to the transition in the orientational order between the coexisting phases.
Hence, the alignment of the particles with the $x$-axis persists a few particle diameters
deeper into the liquid phase $L$ than the layer structure of the $S_A$ phase is maintained --
unlike in the case $\alpha=0$, i.e., \modifiedRed{in which} the smectic layer normal is parallel
to the interface normal, for which the orientational order \modifiedRed{directly} vanishes when
the smectic layers \modifiedRed{disappear} (see Sec.~\ref{sec:results:para}).
\modifiedRed{We note}, that the vanishing of the orientational \modifiedRed{order} significantly after
(\modifiedRed{upon} approaching the interface from the orientational ordered phase) the structural
transition \modifiedRed{associated with} the density profile, has already been observed
previously~\cite{Praetorius_et_al2013} \modifiedRed{in} the case of the interface between an
isotropic liquid and a plastic-triangular crystal (PTC).

For the type of \modifiedRed{ILCs} with the charges at the tips, at low temperatures the new wide
smectic-A phase $S_{AW}$ can be observed (see Fig.~\ref{fig:pd_ilc}(b)). It is characterized by an
alternating structure of layers in which the particles are \modifiedRed{predominantly} parallel to
the layer normal $\vec{\hat n}=\vec{\hat x}$ (like in the $S_A$ phase) and layers of particles
\modifiedRed{which} are preferentially perpendicular to the layer normal.
The free interface formed between the isotropic liquid $L$ and the $S_{AW}$ phase
for $T^*=0.9$ and $\alpha=\pi/2$ is shown in Fig.~\ref{fig:if_ilc_perp_SAW}.
The red regions in Fig.~\ref{fig:if_ilc_perp_SAW}(a) show the layers of particles
(at $x=0$ and $x/R\approx\pm d/R=\pm7.5$) being parallel to the layer normal,
while in between (at $x/R\approx\pm d/(2R)=\pm3.75$) in light blue color the secondary layers are
visible. The dark blue color at $x/R\approx\pm d/(2R)=\pm3.75$ in panel (b) shows that
the orientational order parameter $S_2(x,z)$ \modifiedRed{is negative} at the location of the
secondary layers, because \modifiedRed{there} the particles are preferentially perpendicular
to the layer normal. The interface at the parallel layers behaves very much like the $L$-$S_A$
interface, as can be inferred from the (white) interface contours
$\tilde z_\eta(x/R=0,\pm7.5)/R\approx0.81$ and $\tilde z_{S_2}(x/R=0,\pm7.5)/R\approx-1.83$
which show that the orientational ordering of the $S_{AW}$ phase persists into
the liquid phase $L$ for a few particle diameters.
This is also apparent from the interface positions $z_\eta/R\approx1.0$ and $z_{S_2}/R\approx-2.3$,
depicted by the black dashed lines in Fig.~\ref{fig:if_ilc_perp_SAW}.
Conversely, at lateral positions $x/R\approx d/(2R)=\pm3.75$ associated \modifiedRed{with} the centers
of the intermediate layers, it turns out that the orientational order \modifiedRed{undergoes the
transition} before the layer structure vanishes if one approaches the interface from the $S_{AW}$ side
($\tilde z_{S_2}(x/R=\pm3.75)/R\approx3.39$ and $\tilde z_\eta(x/R=\pm3.75)/R\approx-0.34$;
in order to guide the eye the magenta dots (\textcolor{Magenta}{$\bullet$})
in Fig.~\ref{fig:if_ilc_perp_SAW} mark these positions). \modifiedGreen{This behavior} is
\modifiedRed{opposite to the above one and is} presumably related to the fact, that the secondary
layers consist of particles \modifiedRed{being preferentially} perpendicular to the layer normal;
unlike the particles in the main layers of the $S_{AW}$ phase or the particles in the $S_A$ layers,
these particles do not align with the layer normal $\vec{\hat n}=\vec{\hat x}$.
Instead they are avoiding an orientation parallel to it.
While the transition across the $L$-$S_A$ interface -- from alignment with the layer normal towards
an isotropic orientational distribution -- results in an \textit{increase} \modifiedRed{of} the
effective particle diameter in the $y$- and $z$-direction, for the secondary $S_{AW}$ layers 
the effective diameter is \textit{decreased} from the $S_{AW}$ phase towards the isotropic liquid $L$.
\modifiedRed{In Fig.~\ref{fig:if_ilc_perp_SAW} there are discontinuities in the (white) interface
contour lines $\tilde z_\eta(x)$ and $\tilde z_{S_2}(x)$, as in Fig.~\ref{fig:if_ilc_perp_SA}.
These discontinuities occur at lateral positions} $\check{x}$ at which the packing fraction
$\eta(\check{x},z\rightarrow\pm\infty)$ or the orientational order parameter
$S_2(\check{x},z\rightarrow\pm\infty)$ take the same value in the isotropic bulk, i.e.,
for $z\rightarrow-\infty$, \modifiedRed{as} in the $S_{AW}$ bulk, i.e., for $z\rightarrow\infty$.

\subsection{\label{sec:results:asymp}Asymptotic behavior}
\begin{figure}[!t]
 \includegraphics[width=0.45\textwidth]{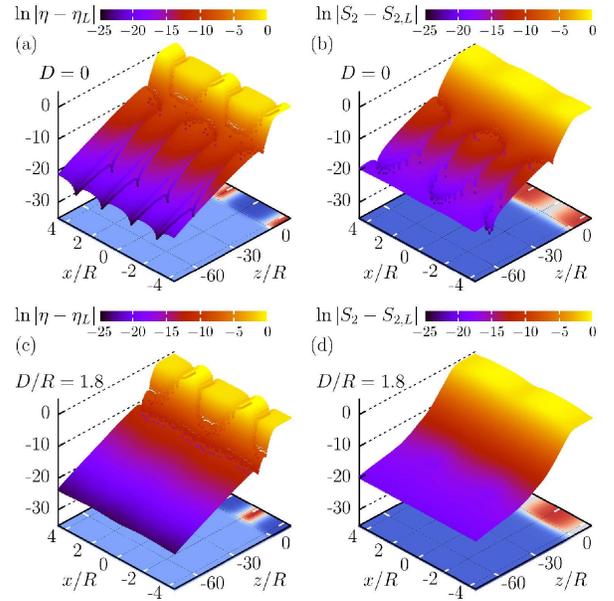}
  \caption{
  $L$-$S_A$ interface profiles of $\eta(x,z)$ and $S_2(x,z)$ for $T^*=10$ and $\alpha=\pi/2$.
  \modifiedRed{Accordingly}, the smectic layer normal $\vec{\hat n}=\vec{\hat x}$ and the interface
  normal (parallel to the $z$-axis) are perpendicular. Panels (a) and (b) show the logarithmic
  deviations $\ln|\eta(x,z)-\eta_L|$ and $\ln|S_2(x,z)-S_{2,L}|$ of the packing fraction
  and the orientational order parameter from their bulk values in the isotropic liquid $L$
  for an \modifiedRed{ILC} with \modifiedRed{the} charges \modifiedRed{concentrated at the center of}
  the molecule, i.e., for $D=0$. Panels (c) and (d) show $\ln|\eta(x,z)-\eta_L|$ and
  $\ln|S_2(x,z)-S_{2,L}|$ for an \modifiedRed{ILC} with \modifiedRed{the} charges at the tips, i.e.,
  for $D/R=1.8$. Note that on the base of each plot the interface profiles $\eta(x,z)$ and $S_2(x,z)$
  are shown in order to elucidate the \modifiedGreen{viewing} angle on the interface.
  \modifiedGreen{The local height of the manifold above the base corresponds to the given color code.}
  Interestingly, for $D=0$ the   periodic structure is still apparent \modifiedRed{even} far away
  from the $L$-$S_A$-interface, i.e., $z/R<-20$, unlike the case $D/R=1.8$, for which the profiles
  are rather flat in lateral direction $x$.
  This can be related to the strong localization of charges \modifiedRed{at} the
  centers of the smectic layers for $D=0$, pronouncing the periodic structure, while for $D/R=1.8$
  the charge sites are spread and less localized along the $x$-direction.
  }
 \label{fig:if_ilc_perp_asymp_liq}
\end{figure}
\begin{figure}[!t]
 \includegraphics[width=0.42\textwidth]{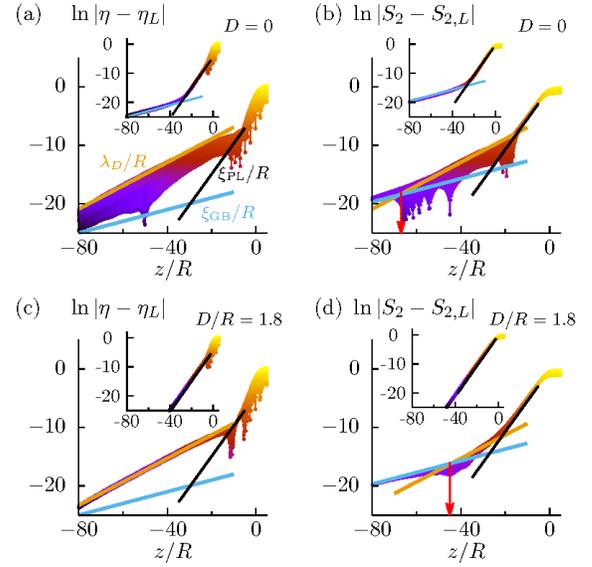}
  \caption{
  The same \modifiedGreen{quantities as shown in} Fig.~\ref{fig:if_ilc_perp_asymp_liq}.
  Panels (a) and (b) correspond to the case $D=0$ presenting $\ln|\eta(x,z)-\eta_L|$ and
  $\ln|S_2(x,z)-S_{2,L}|$, respectively, \modifiedGreen{whereas panels (c) and (d) correspond to
  the case $D/R=1.8$.}
  \modifiedGreen{However, here} the direction of view is parallel to the $x$-axis
  \modifiedGreen{so that the manifold from Fig.~\ref{fig:if_ilc_perp_asymp_liq} is projected
  onto the plane spanned by the vertical axis and the $z$ axis}.
  Away from the interface, i.e., \modifiedRed{for} $z/R<-20$, the decay length for
  $\ln|\eta(x,z)-\eta_L|$ can be identified as the Debye screening length $\lambda_D/R=5$
  for both cases (a) $D=0$ and (c) $D/R=1.8$. From the inset in panel (a), \modifiedRed{which shows}
  $\ln|\eta(x,z)-\eta_L|$ for the corresponding (uncharged) ordinary liquid crystal with
  $L/R=4$ and $\epsilon_R/\epsilon_L=2$, it is apparent that the contributions due to the
  Gay-Berne potential (the asymptotics of which is \modifiedGreen{indicated} by the blue line) and
  \modifiedRed{due to} the hard-core interaction (the asymptotics of which is depicted by the black line)
  are much weaker than the (screened) electrostatic contribution and do not play a role
  \modifiedRed{within the} range of $\ln|\eta(x,z)-\eta_L|$ \modifiedRed{considered here}.
  \modifiedBlue{(In order to guide the eye, the blue and black lines are also shown in the main plots.
  Apparently, in (a) and (c) the blue and black lines are far below the respective profiles.)}
  However, for $\ln|S_2(x,z)-S_{2,L}|$, i.e., \modifiedRed{for} panels (b) and (d), one observes
  \modifiedGreen{crossovers} -- indicated by the intersection of the \modifiedBlue{orange} and blue lines
  at \modifiedRed{$z/R\approx-67$} \modifiedGreen{in (b)} and $z/R\approx-45$ \modifiedGreen{in (d)}
  (compare the red arrows in the respective plots) -- from the electrostatic regime towards the decay
  governed by the Gay-Berne contribution with decay length $\xi_\text{GB}/R\approx10$.
  \modifiedGreen{Such crossovers occur}
  \modifiedBlue{within the considered range $z/R\in[-80,0]$}, because for the orientational
  order parameter the amplitude of the decay, due to the Gay-Berne interaction, is larger than for 
  the packing fraction \modifiedBlue{(compare the intersections of the blue lines with the ordinates
  in panels (a) and (b))}.
  \modifiedRed{Due to the hard-core interaction,} for $z/R>-20$ the decay length
  $\xi_\text{PL}/R\approx1.9$ (\modifiedRed{Parsons-Lee}, black lines) is \modifiedRed{visible}
  for the ordinary liquid crystal in the insets of (a) and (b) as well as for
  $\ln|S_2(x,z)-S_{2,L}|$ of the two considered \modifiedRed{ILCs}.
  (Due to the small amplitudes of the hard-core contributions \modifiedRed{to} $\ln|\eta(x,z)-\eta_L|$,
  for the \modifiedRed{ILC considered here}, this decay \modifiedRed{has not been observed}.)
  In order to confirm, that the decay length $\xi_\text{PL}/R\approx1.9$ is indeed due to the hard-core
  interaction, the insets of \modifiedRed{the} panels (c) and (d) show $\ln|\eta(x,z)-\eta_L|$
  and $\ln|S_2(x,z)-S_{2,L}|$ of the pure hard-core system ($\beta\psi:=\beta\psi_\text{PL}$).
  Interestingly, $\ln|\eta(x,z)-\eta_L|$ and $\ln|S_2(x,z)-S_{2,L}|$ \modifiedGreen{behave} very
  \modifiedGreen{similarly} close to the interface, i.e., $z/R>-10$, for all three kinds of systems
  studied here.
  This suggests that the structure and \modifiedRed{the} orientational properties close to the interface
  are governed by the hard-core interaction which enters into the present DFT approach
  (see Secs.~\ref{sec:theory:model} and \ref{sec:theory:DFT}).
  }
 \label{fig:if_ilc_perp_asymp_liq_plane}
\end{figure}

In this section \modifiedRed{we discuss how} the interface profiles of the packing fraction
$\eta(\vec{r})$ and the orientational order parameter $S_2(\vec{r})$ \modifiedRed{attain} their
respective values $\eta_{L}$ and $S_{2,L}$ in the bulk \modifiedRed{liquid} $L$.
In Fig.~\ref{fig:if_ilc_perp_asymp_liq} the asymptotic behavior is discussed
in terms of $\ln|\eta(x,z)-\eta_L|$ and $\ln|S_2(x,z)-S_{2,L}|$ for $\alpha=\pi/2$ and $T^*=10$,
considering \modifiedRed{ILCs} with charges in the center, i.e.,
$D=0$ (panels (a) and (b)), and with charges at the tips, i.e., $D/R=1.8$ (panels (c) and (d)).
In order to elucidate the view angle on these 3-dimensional logarithmic plots,
\modifiedRed{the interface profiles $\eta(x,z)$ and $S_{2}(x,z)$ are shown in addition
as contour plots (see Fig.~\ref{fig:if_ilc_perp_SA}) at the base of the respective plot.}

Interestingly, while for $D=0$ the periodic structure of the profiles $\eta(x,z)$ and $S_2(x,z)$
in $x$-direction is \modifiedRed{clearly apparent also} in the decays $\ln|\eta(x,z)-\eta_L|$
and $\ln|S_2(x,z)-S_{2,L}|$ far away from the $L$-$S_A$-interface ($z/R<-20$ in
Figs.~\ref{fig:if_ilc_perp_asymp_liq}(a) and (b)), for $D/R=1.8$ (panels (c) and (d))
the decays vary only little as function of $x$.
This distinct behavior can be a signature of the respective molecular charge distributions,
because if the charges are \modifiedRed{localized at} the centers of the molecules, due to the
layer structure in the $S_A$ phase \modifiedRed{the} charges are also localized \modifiedRed{at}
the centers of the smectic layers, while for $D/R=1.8$ the charges are less localized along the
lateral direction $x$. Close to the interface ($z/R>-20$) the structure is very similar
\modifiedRed{in} both cases and, as will be discussed later, it is the hard-core repulsion
which is the dominant contribution here.

Turning the view parallel to the $x$-axis, one obtains projected representations of the logarithmic
plots in Fig.~\ref{fig:if_ilc_perp_asymp_liq}, which are shown in
Fig.~\ref{fig:if_ilc_perp_asymp_liq_plane} keeping the order of panels \modifiedRed{as} in
Fig.~\ref{fig:if_ilc_perp_asymp_liq}.
\modifiedBlue{Hence, Figs.~\ref{fig:if_ilc_perp_asymp_liq_plane}(a) and (b) correspond to the case
$D=0$ presenting $\ln|\eta(x,z)-\eta_L|$ and $\ln|S_2(x,z)-S_{2,L}|$, respectively.
Similiarly, Figs.~\ref{fig:if_ilc_perp_asymp_liq_plane} (c) and (d) show the case $D/R=1.8$.}
\modifiedBlue{In both cases, at large distances, i.e., $z/R<-20$, the decay of the density profiles
is dominated by the electrostatic contribution $U_\text{es}$ to the total interaction potential $U$
(see \modifiedGreen{Figs.}~\ref{fig:if_ilc_perp_asymp_liq_plane}(a) and (c)). Accordingly,}
\modifiedRed{the decay of the envelope} is determined by the Debye screening length $\lambda_D/R=5$,
highlighted by the \modifiedBlue{orange} lines in Fig.~\ref{fig:if_ilc_perp_asymp_liq_plane}.
It is worth mentioning that a DFT study~\cite{Lu_Evans_DaGama1985} of the asymptotic behavior of the
liquid-vapor interface \modifiedRed{has yielded, unlike the present findings,} a decay length $l_b$
larger than the Debye screening length $\lambda_D$ for a hard sphere system with additional Yukawa
interaction. While in the present study the Yukawa potential is purely repulsive,
in Ref.~\cite{Lu_Evans_DaGama1985} \modifiedRed{using} an attractive Yukawa potential is indispensable,
because a sufficiently strong attraction is needed for liquid-vapor coexistence \modifiedRed{to occur}.

Interestingly, the \modifiedRed{asymptotic behavior} of the orientational order parameter at far distances,
i.e., \modifiedRed{for} $z/R<-60$, \modifiedRed{differs} from the electrostatic decay and another
regime (\modifiedBlue{highlighted by blue lines in Fig.~\ref{fig:if_ilc_perp_asymp_liq_plane}})
with \modifiedRed{a} larger decay length $\xi_\text{GB}/R\approx10$ sets in.
This longer-ranged decay is due to the Gay-Berne interaction $U_\text{GB}$ \modifiedRed{which}
is verified by calculating the interface profile for an ordinary liquid crystal \modifiedBlue{(OLC)}
without charges (compare the insets of panels (a) and (b) of Fig.~\ref{fig:if_ilc_perp_asymp_liq_plane}).
\modifiedBlue{For the OLC, at far distances, i.e., $z/R<-30$, the same large decay length
$\xi_\text{GB}/R\approx10$ is observed. However, the amplitudes of the decay of the packing fraction
and \modifiedGreen{of} the orientational order parameter differ significantly.
(The blue line in panel (a) intersects the ordinate at $\ln|\eta-\eta_L|\approx-25$,
whereas the blue line in (b) intersects the ordinate at $\ln|S_2-S_{2,L}|\approx-20$.)}
\modifiedBlue{For $D=0$,} it turns out that \modifiedRed{for the orientational order parameter}
the \modifiedGreen{crossover} from the electrostatic decay towards the Gay-Berne decay
\modifiedRed{occurs at $z/R\approx-67$} \modifiedBlue{(this position is marked by the red arrow
in Fig.~\ref{fig:if_ilc_perp_asymp_liq_plane}(b)), whereas for the case
$D/R=1.8$ the \modifiedGreen{crossover} occurs at $z/R\approx-45$ (see the red arrow in
Fig.~\ref{fig:if_ilc_perp_asymp_liq_plane}(d))}.
Ultimately, the larger Gay-Berne decay length $\xi_\text{GB}/R\approx10$ will also \modifiedRed{become}
apparent in the decay profile of the packing fraction. However, due to the 
\modifiedBlue{smaller amplitude of the Gay-Berne decay of the density compared \modifiedGreen{with}
the decay of the orientational order parameter (compare the insets in
Figs.~\ref{fig:if_ilc_perp_asymp_liq_plane}(a) and (b))}, \modifiedRed{in \modifiedGreen{the present}
case the \modifiedGreen{crossover} occurs} further away from the interface
(in Fig.~\ref{fig:if_ilc_perp_asymp_liq_plane}(a) the intersection of the
\modifiedBlue{orange} line and the blue line \modifiedBlue{is located at $z/R\approx-121$}
\modifiedRed{(not visible)} and in \modifiedRed{Fig.~\ref{fig:if_ilc_perp_asymp_liq_plane}(c)} at
\modifiedBlue{$z/R\approx-97$} \modifiedBlue{(also not visible)}).
\modifiedBlue{However, at very far distances $z/R<-80$, the magnitudes $\ln|\eta-\eta_L|\lesssim-25$ are
very small and cannot be resolved numerically. For this reason,
\modifiedGreen{in Figs.~\ref{fig:if_ilc_perp_asymp_liq_plane}(a) and (c) crossovers} from the electrostatic
regime to the Gay-Berne regime are not shown.}

\modifiedGreen{We note} that, although the Gay-Berne potential $U_\text{GB}$ decays \modifiedRed{algebraicly}
$\propto (r_{12}/R)^{-6}$ (see Eq.~(\ref{eq:Pairpot_GB})), here the Gay-Berne decay is exponential,
because solving the Euler-Lagrange equation \modifiedRed{in} Eq.~(\ref{eq:ELG}) requires the
evaluation of the ERPA contribution $\beta\psi_\text{ERPA}$ of the effective one particle potential
$\beta\psi$ (see Eqs.~(\ref{eq:Eff1Pot_ERPA}) and (\ref{eq:Calc_c1_Approx_interface})).
The numerical calculation of this integral (which extends over the whole volume $\mathcal{V}$
\modifiedGreen{of the system}) requires a truncation in terms of a cut-off distance of the integral
which leads to an exponential decay of this contribution, instead of a power law decay
\modifiedGreen{$\propto (z/R)^{-3}$~\cite{DeGennes1981,Barker1982,Lu_Evans_DaGama1985}},
as it \modifiedGreen{is} expected for the full Gay-Berne potential $U_\text{GB}$.
(The exponent $3$ arises because the asymptotic behavior of an interfacial density profile,
generated by long-ranged forces, varies proportional to the corresponding (total) potential,
\clearpage
which acts on a test particle at a distance $z$ from the interface and which is due to
the pair interaction between the particles in one of the two coexisting phases
(which are separated by the considered interface) and the test particle. Thus, via an integration of the
Gay-Berne pair interaction, which decays $\propto (r_{12}/R)^{-6}$, over a half-space,
one obtains the corresponding total potential decaying
$\propto (z/R)^{-3}$~\cite{DeGennes1981,Barker1982,Dietrich1991}.)

\MODIFIED{\MODIFIEDtwo{For $z/R\rightarrow-\infty$ the algebraic} decay of the Gay-Berne
\MODIFIEDtwo{interaction} potential always dominates the exponential decay due
to the \MODIFIEDtwo{screened} electrostatic interaction, \MODIFIEDtwo{independent} of the relative
strength of the electrostatic and the Gay-Berne interaction potential.
A \MODIFIEDtwo{variation of their} relative strength $\gamma/(R\epsilon_0)$
would only lead to a \MODIFIEDtwo{shift of} the location of the \MODIFIEDtwo{corresponding} crossovers
in the density and \MODIFIEDtwo{the} order parameter profiles (\MODIFIEDtwo{see} the red arrows in
Fig.~\ref{fig:if_ilc_perp_asymp_liq_plane}) \MODIFIEDtwo{caused} by altering the amplitudes of the
respective decays \MODIFIEDtwo{of the two interactions}.}

Close to the interface, i.e., \modifiedRed{for} $-20<z/R<-5$, \modifiedRed{in the insets of
Fig.~\ref{fig:if_ilc_perp_asymp_liq_plane} one can observe} an exponential decay with a decay
length $\xi_\text{PL}/R\approx1.9$ (depicted by the black lines) which arises from the pure
hard-core Parsons-Lee contribution $\beta\psi_\text{PL}$. Thus $\xi_\text{PL}$ can be identified
as the isotropic-liquid bulk correlation length of the pure hard-core system.
Interestingly, while the hard-core correlation length $\xi_\text{PL}$ is observable
in \modifiedRed{OLCs -- within \modifiedGreen{both}} the $\eta$ and \modifiedGreen{the} $S_2$ profiles
(\modifiedBlue{at distances $z/R\in[-20,-5]$ the respective decays closely follow the black lines
which depict the hard-core decay in the insets of Figs.~\ref{fig:if_ilc_perp_asymp_liq_plane}(a) and (b)})
\modifiedRed{--, for ILCs} this decay is visible \modifiedGreen{only} \modifiedRed{within} the $S_2$ profile.
Only for the $S_2$ profile the amplitude of the hard-core decay is large enough, such that the
hard-core correlation length $\xi_\text{PL}$ is observable before the electrostatic decay becomes
\modifiedRed{dominant}. The insets in \modifiedGreen{Figs.}~\ref{fig:if_ilc_perp_asymp_liq_plane}(c) and (d)
show the interface profiles calculated for the pure hard-core system ($\beta\psi:=\beta\psi_\text{PL}$)
in order to verify that the decay close to the interface, i.e., \modifiedRed{for} $-20<z/R<-5$,
is governed by the hard-core interaction.

Finally, it is worth mentioning that \modifiedRed{for all cases shown in
Fig.~\ref{fig:if_ilc_perp_asymp_liq_plane},} the \modifiedRed{structural} and orientational properties
close to the interface, i.e., \modifiedRed{for} $z/R>-10$, agree very well. Thus,
it is the hard-core interaction which determines the structural and orientational 
properties close to the interface, while the electrostatic and the Gay-Berne contributions
\modifiedBlue{dominate further away from the interface}. At intermediate distances electrostatics
dominates the decay of the interface profiles \modifiedRed{whereas} far away from the interface
ultimately the attractive Gay-Berne interaction \modifiedRed{dominates}. Furthermore,
the \modifiedRed{positions} of \modifiedGreen{the crossovers} between these regimes are distinct for the 
packing fraction profile and the orientational order parameter profile.

\subsection{\label{sec:results:tilt}Tilted interfaces}
\begin{figure}[!t]
 \includegraphics[width=0.45\textwidth]{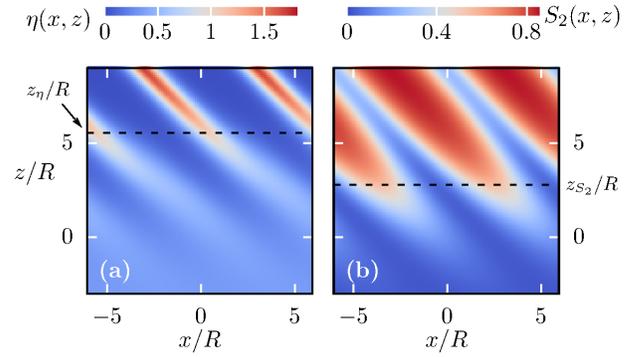}
  \caption{
  The $L$-$S_A$ interface profiles $\eta(x,z)$ \modifiedGreen{(Eq.~(\ref{eq:numberdensity}))} and $S_2(x,z)$
  \modifiedGreen{(Eq.~(\ref{eq:oriorderparameter}))} for $\alpha=\pi/4$
  and $T^*=1.3$ are shown. Here, an \modifiedRed{ILC} with charges \modifiedRed{localized at its}
  center is considered  ($L/R=4,\epsilon_R/\epsilon_L=2,\gamma/(R\epsilon_0)=0.045,\lambda_D/R=5$,
  and $D=0$). For $z\rightarrow-\infty$ the isotropic liquid bulk $L$ is approached and for
  $z\rightarrow\infty$ the bulk of the $S_A$ phase \modifiedRed{is attained}, i.e., the interface
  normal is parallel to the $z$-axis. The red stripes at the top of the contour plots show the tails
  of the smectic layers. The black dashed lines mark the interface positions $z_\eta/R\approx5.56$
  and $z_{S_2}/R\approx2.79$ calculated via Eqs.~(\ref{eq:gibbs_dividing_surface_evaluation_eta}) and
  (\ref{eq:gibbs_dividing_surface_evaluation_s2}). Similar to the case $\alpha=\pi/2$
  \modifiedRed{(see Fig.~\ref{fig:if_ilc_perp_SA}), to a certain extent the orientational order
  persists} into the liquid phase $L$.
  }
 \label{fig:if_ilc_tilt_SA}
\end{figure}
\begin{figure}[!t]
 \includegraphics[width=0.45\textwidth]{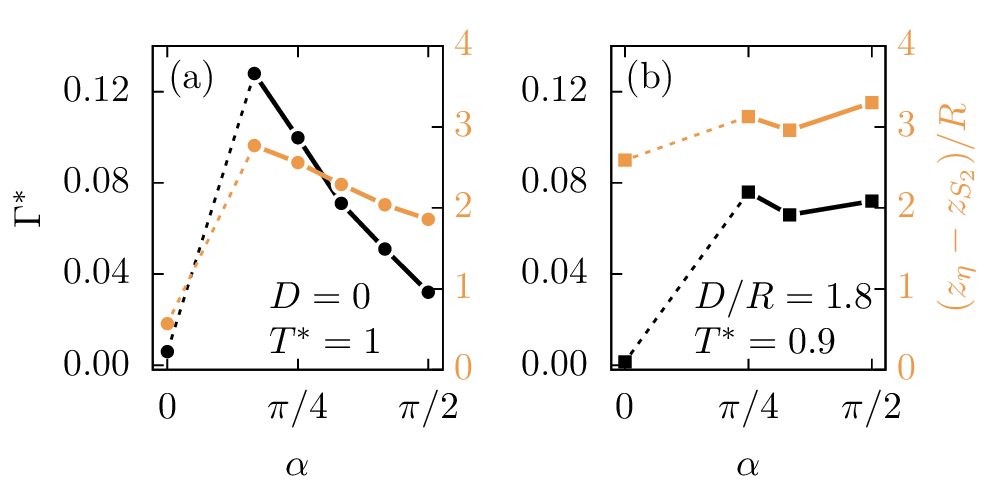}
  \caption{
  The (reduced) interfacial tension $\Gamma^*(\alpha)$ \modifiedRed{(Eq.~(\ref{eq:interface_tension}),
  black line)} and the distance $z_\eta-z_{S_2}$  between the transition in the \modifiedRed{structural}
  and the orientational order \modifiedRed{(orange line)} as function
  of the tilt angle $\alpha$. In panel (a) the $L$-$S_A$ interface at $T^*=1$ is considered
  for \modifiedRed{ILCs} with \modifiedRed{their} charges \modifiedRed{localized at} the center
  \modifiedBlue{($L/R=4,\epsilon_R/\epsilon_L=2,\gamma/(R\epsilon_0)=0.045,\lambda_D/R=5$, and
  $D=0$ (\modifiedGreen{see} Fig.~\ref{fig:pd_ilc}(a))}.
  There are two minima: the global minimum at the equilibrium tilt angle $\alpha_\text{eq}=0$
  (\modifiedRed{i.e.,} interface normal and smectic layer normal are parallel) and a local minimum at
  $\alpha=\pi/2$ which shows that the \modifiedRed{orthogonal} orientation of the smectic layer normal
  and the interface normal is \modifiedRed{a metastable configuration}. The increase \modifiedRed{of}
  the interfacial tension \modifiedBlue{below} $\alpha=\pi/2$ is accompanied by an increase
  \modifiedRed{of} the distance $z_\eta-z_{S_2}$. \modifiedRed{This suggests that maintaining
  to a certain extent the local orientational order in the isotropic liquid  beyond the smectic layers
  costs free energy.}
  \modifiedBlue{
  For technical reasons we did not study small tilt angles $\alpha>0$. \modifiedGreen{Hence we} cannot
  comment on the functional form of $\Gamma^*(\alpha)$ for $0<\alpha<\pi/6$ in the case $D/R=0$ or for
  $0<\alpha<\pi/4$ in the case $D/R=1.8$.
  This is \modifiedGreen{indicated} by connecting the data points at $\alpha=0$ and $\pi/6$ by dashed
  lines in (a) (see the discussion in the main text of Sec.~\ref{sec:results:tilt}).}
  In panel (b) the $L$-$S_{AW}$ interface, \modifiedBlue{which is accessible for ILCs with their
  charges at the tips ($L/R=4,\epsilon_R/\epsilon_L=2,\gamma/(R\epsilon_0)=0.045,\lambda_D/R=5$, and
  $D/R=1.8$)}, is considered for $T^*=0.9$ \modifiedRed{(see Fig.~\ref{fig:pd_ilc}(b)).
  Also in this case} the equilibrium tilt angle $\alpha_\text{eq}=0$ corresponds to the parallel
  orientation of the interface normal and \modifiedRed{the} layer normal.
  \modifiedRed{Below} $\alpha=\pi/2$, \modifiedRed{as function of $\alpha$} the interfacial tension
  is rather flat, \modifiedRed{taking the} value $\Gamma^*\approx0.07$.
  Thus, for the $L$-$S_{AW}$ interface the perpendicular orientation of the interface normal and
  \modifiedRed{of} the smectic layer normal \modifiedRed{corresponds to a labile configuration.}
  \modifiedBlue{(Analogously to panel (a), the data points at $\alpha=0$ and $\pi/4$ in (b) are
  connected by a dashed line.)} \modifiedRed{We note that} $\Gamma^*(\alpha)$ is symmetric around
  $\alpha=\pi/2$, due to the mirror-symmetry of the particles.
  }
 \label{fig:if_tension}
\end{figure}
\begin{figure}[!t]
 \includegraphics[width=0.45\textwidth]{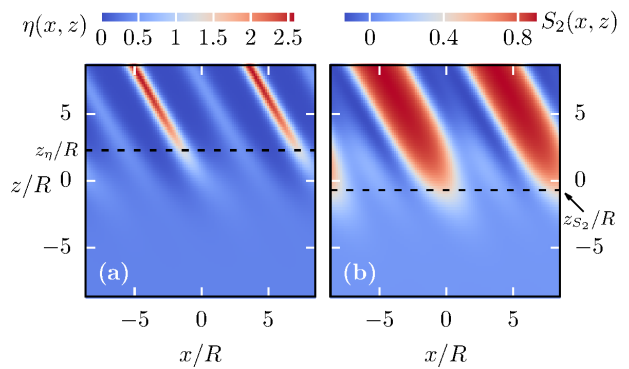}
  \caption{
  Same as Fig.~\ref{fig:if_ilc_tilt_SA}. Here, the $L$-$S_{AW}$ interface profiles $\eta(x,z)$
  \modifiedGreen{(Eq.~(\ref{eq:numberdensity}))} and $S_2(x,z)$
  \modifiedGreen{(Eq.~(\ref{eq:oriorderparameter}))} \modifiedRed{are shown} for $\alpha=\pi/3$
  and $T^*=0.9$. \modifiedRed{To this end}, an ionic liquid crystal with charges at the tips is considered
  ($L/R=4,\epsilon_R/\epsilon_L=2,\gamma/(R\epsilon_0)=0.045,\lambda_D/R=5$, and $D/R=1.8$).
  For $z\rightarrow-\infty$ the isotropic liquid bulk $L$ is approached and for $z\rightarrow\infty$
  the bulk of the $S_{AW}$ phase, i.e., the interface normal is parallel to the $z$-axis.
  The transition in the structure occurs at $z_\eta/R\approx2.28$ and the transition in the
  orientational order \modifiedRed{does so} at $z_{S_2}/R\approx-0.68$.
  }
 \label{fig:if_ilc_tilt_SAW}
\end{figure}

In this section \modifiedRed{we discuss} the dependence of the structural and orientational properties
of the liquid-smectic-interface on the tilt angle $\alpha$. In Fig.~\ref{fig:if_ilc_tilt_SA} the
$L$-$S_A$-interface profiles $\eta(x,z)$ and $S_2(x,z)$ are shown for \modifiedRed{the} reduced
temperature $T^*=1.3$ (see the black dots (\textcolor{black}{$\bullet$}) in Fig.~\ref{fig:pd_ilc}(a))
and $\alpha=\pi/4$. Here, \modifiedRed{we consider} the case of \modifiedRed{ILCs with the charges
localized in the center} ($L/R=4,\epsilon_R/\epsilon_L=2,\gamma/(R\epsilon_0)=0.045,\lambda_D/R=5$,
and $D=0$). Like in the case $\alpha=\pi/2$, \modifiedRed{(see Sec.~\ref{sec:results:perp})} i.e.,
the interface normal and \modifiedRed{the} smectic layer normal $\vec{\hat n}=\vec{\hat x}$ are
perpendicular, a persisting orientational order can be observed at the interface:
The structural transition \modifiedRed{occurs} at $z_\eta/R\approx5.56$, whereas the transition in the
orientational order between the two phases \modifiedRed{takes place} at $z_{S_2}/R\approx2.79$ which
is a few diameters deeper in the isotropic liquid.

In Fig.~\ref{fig:if_tension} the interfacial tension $\Gamma^*(\alpha)$ given by
Eq.~(\ref{eq:interface_tension}) and the distance $z_\eta-z_{S_2}$ between the interface
\modifiedRed{positions} associated with the mean packing fraction $\eta_0(x)$ and the mean
orientational order parameter $S_{20}(x)$ are shown as function of the tilt angle $\alpha$.
In \modifiedRed{Fig.~\ref{fig:if_tension}(a) the case of the $L$-$S_A$-interface for ILCs with the
charges at their center is considered for $T^*=1$}. Both the interfacial tension $\Gamma^*(\alpha)$
(black dots, $\bullet$) and the distance $z_\eta-z_{S_2}$ (orange dots, \textcolor{orange}{$\bullet$})
\modifiedRed{exhibit} a global minimum at $\alpha=0$ and a second, local minimum at $\alpha=\pi/2$.
Thus, the equilibrium tilt angle $\alpha_\text{eq}=0$ corresponds to 
\modifiedRed{the configuration in which} the interface normal and the smectic layer normal
\modifiedRed{$\vec{\hat n}=\vec{\hat z}$ are parallel}, whereas the \modifiedRed{corresponding}
perpendicular orientation $\alpha=\pi/2$ is metastable.
This increase in the interfacial tension $\Gamma^*$ \modifiedRed{below} $\alpha=\pi/2$ suggests that 
the \modifiedRed{configuration, in which} the interface normal and the layer normal 
\modifiedRed{are orthogonal}, should be observable without \modifiedRed{resorting to} any external
stabilizing field which could be \modifiedRed{provided, e.g.,} by a suitably structured substrate.
\modifiedRed{This} metastability of the tilt angle $\alpha=\pi/2$ \modifiedRed{can} be checked
\modifiedRed{also} via computer simulations. Interestingly, the increase \modifiedRed{of} the
interfacial tension \modifiedRed{below} $\alpha=\pi/2$ is accompanied by an increase in the distance
$z_\eta-z_{S_2}$, suggesting that maintaining the local orientational order
\modifiedRed{in the isotropic liquid} beyond the smectic layers costs \modifiedRed{free} energy.
Consistently, in the case $\alpha_\text{eq}=0$, for which the orientational order vanishes
\modifiedRed{directly} with the \modifiedRed{disappearance} of the smectic layers, the 
\modifiedRed{cost in free energy} is lowest. Apparently, \modifiedRed{for $\alpha=0$} the interfacial
tension $\Gamma^*(\alpha=0)\approx0.006$ is significantly smaller \modifiedRed{than for} all other
angles $\alpha$ shown in Fig.~\ref{fig:if_tension}(a).
\modifiedBlue{For technical reasons we did not study small tilt angles $\alpha>0$ and hence cannot
comment on the functional form of $\Gamma^*(\alpha)$ for $0<\alpha<\pi/6$ in the case $D/R=0$
or for $0<\alpha<\pi/4$ in the case $D/R=1.8$.
This is \modifiedGreen{indicated} by connecting the data points at $\alpha=0$ and $\pi/6$ by dashed lines.
(For the same reason, in (b) the data points at $\alpha=0$ and $\pi/4$ are connected by dashed lines.)}
It has been pointed out in Sec.~\ref{sec:theory:DFT}, that due to the \modifiedGreen{crossover}
\modifiedRed{at the tilt angle $\alpha=0$} from a periodic system towards
\modifiedRed{one which is translationally} invariant in lateral direction $x$, the integration domain
$\mathcal{V}_d$ for evaluating the coefficients $Q_i(\vec{r})$
\modifiedRed{(see Eq.~(\ref{eq:ExpansionCoeffs_interface}))} is not continuously evolving 
\modifiedBlue{at $\alpha=0$}.
For $\alpha>0$ it is \modifiedRed{a} slice of length $d_x=d/\sin(\alpha)$ in
$x$-direction, while \modifiedRed{for $\alpha=0$} it is the subsystem of length $d$ in $z$-direction at
position $\vec{r}$. (\modifiedGreen{For} $\alpha=0$ the extent in $x$- and $y$-direction is arbitrary due
to the translational invariance in lateral direction.) In order to describe a continuous 
\modifiedRed{variation of} the interfacial tension $\Gamma^*(\alpha)$ for all tilt angles
$\alpha\in[0,\pi/2]$, one \modifiedGreen{thus} needs to consider a different approach,
\modifiedBlue{which does not rely on a projected density and thereby on the direction of the bulk
smectic layer normal $\vec{\hat n}$ throughout the whole \modifiedGreen{interface structure}.}
\modifiedRed{Nonetheless, our above approach still allows one} to compare the interfacial
tension $\Gamma^*(\alpha)$ for the extreme cases $\alpha=0$ and $\pi/2$, \modifiedRed{thus predicting}
which one \modifiedRed{of the two} is preferred.
Furthermore, \modifiedRed{our approach provides an understanding} \modifiedGreen{of} the local increase in 
$\Gamma^*(\alpha)$ \modifiedRed{below} $\alpha=\pi/2$, as \modifiedRed{one observes} an increasing
distance $z_\eta-z_{S_2}$ between the transition in the \modifiedRed{structural} and the orientational
order at the interface.

Figure~\ref{fig:if_tension}(b) shows data for the $L$-$S_{AW}$-interface at $T^*=0.9$ for ILCs with
charges \modifiedRed{located} at the tips. Around $\alpha=\pi/2$ the interfacial tension
(black \modifiedRed{squares}, {\tiny$\blacksquare$}) is a rather flat function of $\alpha$ 
\modifiedRed{taking values around} $\Gamma^*\approx0.07$. The slight variations in
$\Gamma^*$ for $\alpha\in[\pi/4,\pi/2]$ might be \modifiedRed{caused by} the numerical evaluation
of Eq.~(\ref{eq:ELG}) which has to be done separately for each tilt angle $\alpha$.
\modifiedRed{Consistently}, the distance $z_\eta-z_{S_2}$ (orange \modifiedRed{squares},
{\tiny\textcolor{orange}{$\blacksquare$}}) does not vary much \modifiedRed{as function of} the tilt
angle $\alpha$. \modifiedRed{As above}, the equilibrium tilt angle $\alpha_\text{eq}=0$ corresponds
to the \modifiedRed{configuration in which} the interface normal and the smectic layer normal
\modifiedRed{$\vec{\hat n}=\vec{\hat z}$ are parallel}.

Finally, in Fig.~\ref{fig:if_ilc_tilt_SAW}, \modifiedRed{we show} the contour plot of the
$L$-$S_{AW}$-interface for $\alpha=\pi/3$ and $T^*=0.9$ \modifiedGreen{for} an \modifiedRed{ILC} system
with $D/R=1.8$, \modifiedRed{illustrating} the structure of this \modifiedRed{type} of interface. 

\section{\label{sec:summary}Summary and conclusions}

Free interfaces in systems composed of ionic liquid crystals (ILCs) have been studied within
density functional theory (see Sec.~\ref{sec:theory:DFT}). In particular, the discussion
\modifiedRed{has been} focused on two kinds of ionic liquid crystals:
\modifiedRed{first}, ILCs with \modifiedRed{the} charges \modifiedRed{localized at} the center of the
molecules, i.e., $D=0$ (see Figs.~\ref{fig:ellipsoids} and \ref{fig:pairpot}), and, second, ILCs with
\modifiedRed{the} charges at the tips of the molecules, i.e., $D/R=1.8$. All other model parameters,
i.e., $L/R=4,\epsilon_R/\epsilon_L=2,\gamma/(R\epsilon_0)=0.045,\lambda_D/R=5$, are identical in both
cases. \modifiedRed{Therefore} the two kinds differ solely by the charge distribution
\modifiedRed{within the molecules}.

For $D=0$ coexistence between the isotropic liquid $L$ and the ordinary smectic-A phase $S_A$
can be observed at \modifiedRed{a} sufficiently large mean packing fraction $\eta_0$
\modifiedRed{(see Fig.~\ref{fig:pd_ilc}(a))}. The $S_A$ phase is characterized by a layered
structure in the direction of the smectic layer normal $\vec{\hat n}$ with \modifiedRed{a}
smectic layer spacing $d\approx L$ comparable to the particle length $L$.
Within the smectic layers the particles are well aligned with the smectic layer normal.
The phase behavior of ILCs is altered by varying the molecular charge distribution,
as can be \modifiedRed{inferred from comparing the case $D=0$ (i.e., charges at the center) and}
$D/R=1.8$ \modifiedRed{(i.e., charges at the tips, see Fig.~\ref{fig:pd_ilc}(b))}.
At sufficiently low \modifiedRed{temperatures} a new smectic-A phase \modifiedRed{has been} observed,
which is referred to as \modifiedRed{the} $S_{AW}$ phase~\cite{Bartsch2017}.
The $S_{AW}$ phase shows an alternating structure of layers with \modifiedRed{the} majority of
\modifiedRed{the} particles being oriented parallel to the smectic layer normal $\vec{\hat n}$ and 
\modifiedRed{the} minority of \modifiedRed{the} particles localized in secondary layers which prefer
orientations perpendicular to $\vec{\hat n}$. Due to the alternating layer structure, the smectic
layer spacing $d/R\approx7.5$ \modifiedRed{in the $S_{AW}$ phase} is increased
\modifiedRed{compared with the spacing in the $S_A$ phase}. 

For a parallel orientation of the smectic layer normal $\vec{\hat n}=\vec{\hat z}$ and the
$L$-$S_A$-interface normal, i.e., \modifiedRed{for} $\alpha=0$ (see Fig.~\ref{fig:interface_sketch}),
it turns out that the interface locations $z_\eta$ and $z_{S_2}$, associated with the transition in
\modifiedRed{the structural} and in the orientational order, \modifiedRed{respectively}, are very
close to each other (see Fig.~\ref{fig:if_ilc_l-sa_para}). In fact, Fig.~\ref{fig:gibbs_parallel}
shows that  for the whole temperature range considered here, the \modifiedRed{difference}
$z_\eta-z_{S_2}<d$ in the two interface positions is smaller than the smectic layer spacing $d$.
Hence, \modifiedRed{for $\alpha=0$} the orientational order vanishes within the last smectic layer
at the $L$-$S_A$-interface. \modifiedRed{Concerning the interface positions},
Fig.~\ref{fig:gibbs_parallel} demonstrates that \modifiedRed{ILCs} with $D/R=1.8$ and ordinary
(uncharged) liquid crystals with $L/R=4$ and $\epsilon_R/\epsilon_L=2$ exhibit qualitatively the
same results. Considering the $L$-$S_{AW}$-interface \modifiedRed{(see Fig.~\ref{fig:if_ilc_l-saw_para})}
one observes an increase in $z_\eta-z_{S_2}$, but it \modifiedRed{remains significantly smaller} than
the smectic layer spacing $d/R\approx7.5$. Thus, for $\alpha=0$ it turns out that the loss of
orientational order coincides with the \modifiedRed{disappearance} of the layer structure of the
respective smectic-A phase at the interface towards the isotropic liquid.
\modifiedRed{This holds for all parameter values studied here.}

Interestingly, for $\alpha=\pi/2$, i.e., changing the relative orientation of the
smectic layer normal $\vec{\hat n}=\vec{\hat x}$ and the interface normal such that they
are perpendicular to each other, leads to qualitative changes in the interfacial properties:
\modifiedRed{a} periodic structure of the interface in lateral direction $x$ can be observed, which is
a direct consequence of the periodicity in the bulk smectic-A phase with the smectic layer spacing $d$
(see Figs.~\ref{fig:interface_sketch}, \ref{fig:if_ilc_perp_SA}, and \ref{fig:if_ilc_perp_SAW}).
For the $L$-$S_A$-interface \modifiedRed{(see Fig.~\ref{fig:if_ilc_perp_SA})} one observes considerable
\modifiedRed{differences} $(z_\eta-z_{S_2})/R\gtrsim2$ \modifiedRed{between} the interface positions.
Thus, the (nearly) parallel orientations of particles in the $S_A$ layers persists a few particle
diameters $R$ into the liquid phase $L$, unlike the case $\alpha=0$, \modifiedRed{for which} the
orientational order vanishes \modifiedRed{directly} with the breakdown of the $S_A$ layer structure
at the interface, i.e., within the last smectic layer.
Due to the periodicity in (lateral) $x$-direction, in the case $\alpha=\pi/2$ one indeed observes 
a qualitative change in the structure \modifiedRed{of} the $L$-$S_{AW}$-interface compared to the
$L$-$S_A$-interface. While at the tails of the $S_{AW}$ main layers the interface also features an
orientational order \modifiedRed{which} continues further into the liquid phase $L$ than the layer
structure ($(\tilde z_\eta(x)-\tilde z_{S_2}(x))/R\approx2.6$). For the secondary layers it is the 
layer structure that persists deeper into the $L$ phase than the orientational order
($(\tilde z_\eta(x)-\tilde z_{S_2}(x))/R\approx-3.73$). The \modifiedRed{opposite} behavior at the
main, respectively secondary, layers is presumably driven by the orientational properties of the
respective kinds of layers: \modifiedRed{in} the main layers the particles are well aligned with
the smectic layer normal $\vec{\hat n}=\vec{\hat x}$ and therefore show an effective diameter in
the $y$-$z$-plane \modifiedRed{which} is comparable to the particle diameter $R$. However, in the
secondary layers (here \modifiedRed{with} $S_2(x,z)<0$) the particles avoid orientations parallel
to the $x$-axis, giving rise to an considerably larger effective radius.
\modifiedRed{Upon approaching the liquid phase $L$, this effective radius \textit{increases} for
the main layers of the $S_{AW}$ phase, whereas it \textit{decreases} for the secondary
layers.}

In Sec.~\ref{sec:results:asymp} the asymptotic behavior of the interface profiles has been studied.
In particular, in Figs.~\ref{fig:if_ilc_perp_asymp_liq} and \ref{fig:if_ilc_perp_asymp_liq_plane}
the $L$-$S_A$-interface for $\alpha=\pi/2$ \modifiedRed{has been} considered for the two ILC systems
with $D/R=0$ and $1.8$. For $D=0$, i.e., \modifiedRed{with the charges being localized at} the center,
the periodic structure of the interface is apparent from \modifiedRed{the quantities} 
$\ln|\eta(x,z)-\eta_L|$ and $\ln|S_2(x,z)-S_{2,L}|$, showing the logarithmic deviations of the profiles
$\eta(x,z)$ and $S_2(x,z)$ from their respective liquid bulk values $\eta_L$ and $S_{2,L}$
(Figs.~\ref{fig:if_ilc_perp_asymp_liq_plane}(a) and (b)), \modifiedRed{which can be resolved}
even at far distances $z/R<-20$ from the $L$-$S_A$-interface. Conversely, for $D/R=1.8$, i.e.,
\modifiedRed{the charges being fixed} at the tips, \modifiedRed{far from the interface}
$\ln|\eta(x,z)-\eta_L|$ and $\ln|S_2(x,z)-S_{2,L}|$ vary only marginally as function of the lateral
coordinate $x$. While for $D=0$ the charges are strongly localized \modifiedRed{at} the centers of the
smectic layers, \modifiedRed{thus} promoting the periodic structure, for $D/R=1.8$ the charges are less
localized and more distributed along the $x$-direction.

The asymptotic decays \modifiedRed{of the interface profiles} towards the isotropic liquid $L$
show an interesting and rich behavior. \modifiedRed{We have found} three distinct \modifiedRed{spatial}
regimes, \modifiedRed{which are} associated with the three contributions to the underlying pair
potential \modifiedRed{(see Eq.~(\ref{eq:Pairpot}))}. Although the presence of charges is the
distinctive \modifiedRed{feature} of \modifiedRed{ILCs}, the (screened) electrostatic contribution
\modifiedRed{to the interaction (Eq.~(\ref{eq:PairPot_ES})) governs} the asymptotic decay only at
intermediate distances from the interface \modifiedRed{(see Fig.~\ref{fig:if_ilc_perp_asymp_liq_plane})}.
\modifiedRed{In this regime}, the decay length is given by the Debye screening length,
\modifiedRed{here} $\lambda_D/R=5$. Ultimately, it is the attractive Gay-Berne contribution
\modifiedRed{to the interaction (Eq.~(\ref{eq:Pairpot_GB})) which} dominates the \modifiedRed{outermost}
asymptotic behavior; \modifiedRed{for the system studied here} a considerably large decay length
$\xi_\text{GB}/R\approx10$ is observed, which \modifiedRed{is due to} the truncated power law decay
of the GB potential. Close to the interface, the hard-core interaction, \modifiedRed{which leads to the
Parsons-Lee contribution to the DFT expression (Eq.~(\ref{eq:Eff1Pot_PL})), dominates} the profiles
$\eta(x,z)$ and $S_2(x,z)$. The corresponding decay length $\xi_\text{PL}/R\approx1.9$ is comparable to
the particle diameter $R$. \modifiedBlue{This is \modifiedGreen{plausible}, because for the case
considered here the tilt angle is $\alpha=\pi/2$, i.e., the smectic layer normal is perpendicular
to the interface normal, and thus the particles in the $S_A$ layers are oriented preferentially
perpendicular to the interface normal as well}. 
Interestingly, the \modifiedGreen{crossovers} between these three different regimes occur at
distances \modifiedRed{characteristic} for the packing fraction $\eta(x,z)$ and the orientational
order parameter $S_2(x,z)$. While \modifiedRed{for both types of ILCs considered in
Fig.~\ref{fig:if_ilc_perp_asymp_liq_plane}} all three decay lengths $\xi_\text{PL}$, $\xi_\text{GB}$,
and $\lambda_D$ are apparent from $\ln|S_2(x,z)-S_{2,L}|$, \modifiedRed{from} $\ln|\eta(x,z)-\eta_L|$
only the decay length $\lambda_D$ \modifiedRed{can be inferred within} the considered range $z/R>-80$.
This \modifiedRed{situation} is caused by the relative magnitudes of the respective decay amplitudes:
for the packing fraction profile the decay amplitudes due to the Gay-Berne and the hard-core
interaction are too small, compared to the \modifiedRed{corresponding} amplitude due to the
electrostatic \modifiedRed{interaction}, to be observable.

\modifiedRed{Since the structural} and orientational properties directly at the interface
\modifiedRed{position} are determined by the hard-core interaction, i.e., the Parsons-Lee contribution
$\beta\psi_\text{PL}$ (Eq.~(\ref{eq:Eff1Pot_PL})), to the effective one-particle potential $\beta\psi$,
\modifiedRed{close to the interface} the profiles for ordinary liquid crystals (OLCs) and ILCs
\modifiedRed{with} the same length-to-breadth ratio $L/R$ are very similar. In particular, this includes
the interface positions $z_\eta$ and $z_{S_2}$ \modifiedRed{(see Fig.~\ref{fig:gibbs_parallel})}
associated with the transition in the \modifiedRed{structural and orientational order, respectively.}
Nevertheless the asymptotic behavior, as discussed above, is distinct for the different kinds of
particles (hard ellipsoids, OLCs, and ILCs) and shows a rich phenomenology, specifically for ILCs,
due to the cross-overs between the distinct \modifiedRed{spatial} regimes corresponding to the
\modifiedRed{various} contributions to the pair potential. Additionally, the bulk phase behavior is
crucially affected by the type of particles, because only for the ILCs with charges at the tips,
the \modifiedRed{phase $S_{AW}$} is observed.

Finally, the dependence of the structural and orientational properties of liquid-smectic interfaces
on the tilt angle $\alpha$ between the interface normal and the smectic layer normal has been discussed.
For the $L$-$S_A$-interface \modifiedRed{(see Fig.~\ref{fig:if_tension}(a))},
it turns out, that the parallel orientation of the interface normal and \modifiedRed{of the} smectic
layer normal is the \modifiedRed{one \modifiedGreen{in} thermal equilibrium}, i.e., $\alpha_\text{eq}=0$.
The perpendicular orientation $\alpha=\pi/2$ is metastable. Interestingly, the increase in the
interfacial tension \modifiedRed{below} $\alpha=\pi/2$ is accompanied by an increase in the distance
$z_\eta-z_{S_2}$, suggesting that maintaining the local orientational order beyond the smectic layers
\modifiedRed{towards the isotropic liquid costs free energy}. Consistently, in the case
$\alpha_\text{eq}=0$, for which the orientational order vanishes \modifiedRed{directly} with the
\modifiedRed{disappearance} of the smectic layers, \modifiedRed{the cost of free energy for forming
the interface is lowest}. For the $L$-$S_{AW}$-interface \modifiedRed{(see Fig.~\ref{fig:if_tension}(b))}
again the equilibrium tilt angle $\alpha_\text{eq}=0$ corresponds to the parallel orientation of the
interface and smectic layer normal. However, \modifiedRed{in this case, around $\alpha=\pi/2$,}
the interfacial tension $\Gamma^*(\alpha)$ varies only weakly so that here the
perpendicular orientation is labile.
\MODIFIED{Additional contributions to the surface tensions might arise from elastic deformations
of the director field, i.e., spatial variations of the director $\vec{\hat n}:=\vec{\hat n}(\vec{r})$,
or deviations from a rotational-symmetric distribution of particle orientations around the director,
i.e., $f(\vec{r},\vec\omega)\neq f(\vec{r},\vec{\hat n}\cdot\vec\omega)$. These contributions are
neglected by our approach. Elastic effects can be considered \MODIFIEDtwo{through an} explicit dependence of
the free energy functional on the director field $n(\vec{r})$, \MODIFIEDtwo{i.e.,} via an elastic energy
contribution~\cite{DeGennes1974}. Alternatively, giving up the assumption of a rotational
symmetric distribution of orientations around a particular axis (and thereby enforcing a prescribed
homogeneous director field) would also allow \MODIFIEDtwo{one} to study the deformations of the director
field. However, incorporating these effects would lead to a drastic increase \MODIFIEDtwo{of} the
computational effort.}

Lastly, \modifiedGreen{we emphasize} that \modifiedRed{although} here \modifiedRed{we have focused}
solely on free interfaces between coexisting bulk phases of \modifiedRed{ILCs}, the DFT framework in
Sec.~\ref{sec:theory:DFT} can be \modifiedRed{extended} to inhomogeneous systems of ILCs
exposed\modifiedRed{, e.g.,} to external fields or \modifiedRed{ILC-electrolytes} in contact with
an electrode.

\appendix

\section{\label{sec:appendix:UnevenModes}Implications of \modifiedRed{the presence of} \modifiedGreen{odd} Fourier modes in $\bar\rho(\vec{r},\vec\omega)$}
\begin{figure}[!t]
 \includegraphics[width=0.48\textwidth]{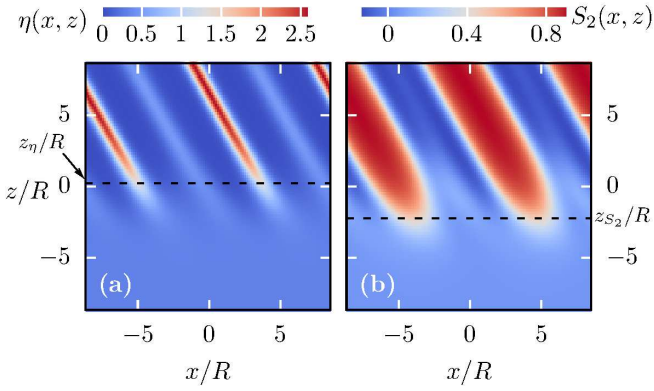}
  \caption{
  Same as Fig.~\ref{fig:if_ilc_tilt_SAW}. Here, the $L$-$S_{AW}$ interface profiles $\eta(x,z)$
  \modifiedGreen{(Eq.~(\ref{eq:numberdensity}))} and $S_2(x,z)$
  \modifiedGreen{(Eq.~(\ref{eq:oriorderparameter}))} \modifiedRed{are calculated} for $\alpha=\pi/3$
  and $T^*=0.9$ \modifiedRed{by} using the projected density containing \modifiedGreen{odd} Fourier-modes
  up to second order.
  The profiles are qualitatively equivalent to those obtained
  without using the \modifiedGreen{odd} modes in the projected density $\bar\rho(\vec{r},\vec\omega)$
  (see Eq.~(\ref{eq:WeightedDensity})). In agreement with the results shown in
  Fig.~\ref{fig:if_ilc_tilt_SAW} one observes \modifiedRed{an} orientational order
  (within the main layers of the $S_{AW}$ phase) \modifiedRed{persisting} up to a few particle
  diameter $R$ into the liquid phase ($(z_\eta-z_{S_2})/R=0.24-(-2.23)=2.47$).
  }
 \label{fig:if_ilc_tilt_SAW_SinTerms}
\end{figure}

In this appendix the implications \modifiedRed{are discussed} of considering the \modifiedRed{the
occurrence of} \modifiedGreen{odd} Fourier modes up to second-order \modifiedRed{ones within}
the projected density $\bar\rho(\vec{r},\vec\omega)$. Including these terms,
$\bar\rho(\vec{r},\vec\omega)$ takes the following modified form:
\begin{equation}
 \begin{split}
  & \bar\rho(\vec{r},\vec\omega,[\rho]) = \frac{1}{4\pi}\bigg[
    Q_0(\vec{r},[\rho])+
    Q_1(\vec{r},[\rho])\cos\left(2\pi(\vec{r}\cdot\vec{\hat n})/d\right)\\
  &+Q_2(\vec{r},[\rho])\cos\left(4\pi(\vec{r}\cdot\vec{\hat n})/d\right)+5P_2(\vec{\omega}\cdot\vec{\hat n})
    \bigg(Q_3(\vec{r},[\rho])\\
  &+Q_4(\vec{r},[\rho])\cos\left(2\pi(\vec{r}\cdot\vec{\hat n})/d\right)
   +Q_5(\vec{r},[\rho])\cos\left(4\pi(\vec{r}\cdot\vec{\hat n})/d\right)\bigg)\\
  &+Q_6(\vec{r},[\rho])\sin\left(2\pi(\vec{r}\cdot\vec{\hat n})/d\right)
   +Q_7(\vec{r},[\rho])\sin\left(4\pi(\vec{r}\cdot\vec{\hat n})/d\right)\\
  &+5P_2(\vec{\omega}\cdot\vec{\hat n})\bigg(
    Q_8(\vec{r},[\rho])\sin\left(2\pi(\vec{r}\cdot\vec{\hat n})/d\right)\\
  &+Q_9(\vec{r},[\rho])\sin\left(4\pi(\vec{r}\cdot\vec{\hat n})/d\right)\bigg)\bigg].
 \end{split}
\label{eq:WeightedDensity_SinTerms}
\end{equation}
This expression differs from Eq.~(\ref{eq:WeightedDensity}) by the (\modifiedGreen{odd} Fourier-)terms
corresponding to the coefficients $Q_i(\vec{r})$ with $i\in[6,\cdots,9]$:
\begin{align}
 Q_i(\vec{r},[\rho])&=\frac{1}{\mathcal{V}_d}
 \Int{\mathcal{V}}{3}{r'}\Int{\mathcal{S}}{2}{\omega'}
 \rho(\vec{r}',\vec\omega')w_i(\vec{r},\vec{r}',\vec\omega'),
 \label{eq:ExpansionCoeffs_SinTerms}
\end{align}
\modifiedRed{where}
\begin{align}
 w_6&=2\mathcal{T}(\vec{r}-\vec{r}')\sin\left(2\pi(\vec{r}'\cdot\vec{\hat n})/d\right),\nonumber\\
 w_7&=2\mathcal{T}(\vec{r}-\vec{r}')\sin\left(4\pi(\vec{r}'\cdot\vec{\hat n})/d\right),\nonumber\\
 w_8&=2\mathcal{T}(\vec{r}-\vec{r}')P_2(\vec\omega'\cdot\vec{\hat n})\sin\left(2\pi(\vec{r}'\cdot\vec{\hat n})/d\right),\nonumber\\
 w_9&=2\mathcal{T}(\vec{r}-\vec{r}')P_2(\vec\omega'\cdot\vec{\hat n})\sin\left(4\pi(\vec{r}'\cdot\vec{\hat n})/d\right).
 \label{eq:ExpansionCoeffs_SinTerms_kernels}
\end{align}
\modifiedRed{The} coefficients $Q_i$ with $i=6,\cdots,9$ vanish for the considered bulk phases,
because smectic-A phases \modifiedRed{exhibit} mirror-symmetry with respect to the layer center.
\modifiedRed{In general,} at interfaces they do not vanish. In order to compare the 
\modifiedRed{corresponding} interface profiles \modifiedRed{
$\eta(\vec{r})=\frac{\pi}{6}LR^2\Int{\mathcal{S}}{2}{\omega}\rho(\vec{r},\vec\omega)$ and
$S_2(\vec{r})=\Int{\mathcal{S}}{2}{\omega}f(\vec{r},\vec\omega)P_2(\vec\omega\cdot\vec{\hat n})$}
\modifiedBlue{(\modifiedGreen{see} Eqs.~(\ref{eq:numberdensity}) and (\ref{eq:oriorderparameter}),
respectively\modifiedGreen{; $\mathcal{S}$ is the full solid angle})}
\modifiedRed{obtained from} solving the Euler-Lagrange equation, i.e., Eq.~(\ref{eq:ELG}),
\modifiedRed{by} using the projected density without the \modifiedGreen{odd} terms given by
Eq.~(\ref{eq:WeightedDensity}) and the projected density containing these terms, i.e.,
\modifiedRed{by} using Eq.~(\ref{eq:ExpansionCoeffs_SinTerms}), the case $\alpha=\pi/3$
\modifiedRed{(see Eq.~(\ref{eq:layernormal}))} and the $L$-$S_{AW}$-interface shown in
Fig.~\ref{fig:if_ilc_tilt_SAW} \modifiedRed{are} considered \modifiedRed{again}.
In Fig.~\ref{fig:if_ilc_tilt_SAW_SinTerms} the \modifiedRed{two} respective profiles are shown
\modifiedRed{by} using Eq.~(\ref{eq:WeightedDensity_SinTerms}): \modifiedRed{for both $\eta(\vec{r})$
and $S_2(\vec{r})$} there are no qualitative differences compared \modifiedRed{with}
Fig.~\ref{fig:if_ilc_tilt_SAW}. The interface positions $z_\eta/R\approx0.24$ and
$z_{S_2}/R\approx-2.23$ are shifted in $z$-direction compared to the results shown in
Fig.~\ref{fig:if_ilc_tilt_SAW}. \modifiedRed{But} their distance $(z_\eta-z_{S_2})/R\approx2.47$
is comparable to the previous results ($(z_\eta-z_{S_2})/R\approx2.96$ in
Fig.~\ref{fig:if_ilc_tilt_SAW}). \modifiedBlue{Hence, in qualitative agreement with the results
shown in Fig.~\ref{fig:if_ilc_tilt_SAW}} one observes a persisting orientational order
(within the main layers of the $S_{AW}$ phase) up to a few particle diameters $R$ into the liquid phase.

\bibliography{literature}
\renewcommand{\baselinestretch}{1.6}

\end{document}